\documentclass{article}

\usepackage{PRIMEarxiv}

\usepackage[utf8]{inputenc} 
\usepackage[T1]{fontenc}    
\usepackage{hyperref}       
\usepackage{url}            
\usepackage{booktabs}       
\usepackage{amsfonts}       
\usepackage{nicefrac}       
\usepackage{microtype}      
\usepackage{lipsum}
\usepackage{fancyhdr}       
\usepackage{graphicx}       
\usepackage{tcolorbox}
\usepackage{amssymb}
\usepackage{amsmath,amssymb,amsfonts}
\usepackage{algorithmic}

\graphicspath{{media/}}     

\title{ORIGAMI: A flexible state channels design for public blockchain systems
}

\author{
  Negka Lydia \\
  Department of Computer Science\\ and Biomedical Informatics \\
  University of Thessaly \\
  Lamia, Greece\\
  \texttt{lnegka@uth.gr} \\
   \And
  Katsika Angeliki \\
  Department of Computer Science\\ and Biomedical Informatics \\
  University of Thessaly \\
  Lamia, Greece\\
  \texttt{akatsika@uth.gr} \\
  \And
  Spathoulas Georgios \\
  Department of Information Security\\ and Communication Technology \\
  NTNU \\
  Gjovik, Norway\\
  \texttt{georgios.spathoulas@ntnu.no} \\
  \And
  Plagianakos Vassilis \\
  Department of Computer Science\\ and Biomedical Informatics \\
  University of Thessaly \\
  Lamia, Greece\\
  \texttt{vpp@uth.gr} \\
}

\begin{document}
\maketitle

\begin{abstract}
Public blockchain systems offer security guarantees that cannot be matched by any centralised system. This offering has attracted a lot of interest and has exposed a significant limitation of most blockchain designs with regards to scalability. One of the scaling solutions proposed is state channels which enables serving given applications with minimum number of transactions. Existing state channels designs set multiple compatibility requirements for applications to be deployed. Origami is a novel state channels design which removes most of the requirements of existing approaches, while it also offers a number of new features. Origami enables dynamic groups of users to interact in an unordered way completely off-chain after an initial on-boarding on-chain transaction. The proposed design is analysed in detail and compared to existing schemes, while a formal security analysis validates the security properties it offers.
\end{abstract}


\section{Introduction}
\label{sec:intro}

Blockchain technology's wide appeal is rooted in the unique security guarantees it offers in comparison with other existing computing approaches. Said guarantees have resulted in very high traffic in popular blockchain networks \cite{treiblmaier2020blockchain}. With Bitcoin \cite{nakamoto2019bitcoin} having an average transaction rate of 7 tx/sec and Ethereum \cite{buterin2014next} in only a slightly better position with 30 tx/sec, most blockchain networks are ill equipped to handle the demanded scale of usage \cite{khan2021systematic}. Serious drawbacks have been brought on as a result, in the form of exorbitant delays and costs that outweigh the benefits of blockchain technology and render it unusable. 

The fault for these limitations lies largely with the Proof of Work (PoW) consensus mechanism that most first generation blockchain networks use \cite{gervais2016security}. A main requirement in PoW is for all transactions to be processed by all nodes before they are published. Additionally, restrictions in block size and block production rate, enforced for security reasons, also contribute to the scalability problem. Even with the adoption of alternative consensus mechanisms such as Proof of Stake (PoS) \cite{saleh2021blockchain} the limitations regarding network capacity have not been diminished. It is evident that in order for blockchain networks to serve as global computing platforms their capacity shall be further extended.

Great resources have been devoted for the purpose of improving blockchain networks' scalability. The emerging solutions are mostly divided into two categories described as Layer 1 and Layer 2 approaches \cite{zhou2020solutions}. Layer 1 approaches target the fundamental design and functionality of the network. Examples include altering the consensus mechanism \cite{bach2018comparative} or incorporating sharding \cite{chow2018sharding,kokoris2018omniledger}. Implementing Layer 1 solutions is a large scale, high cost endeavour in most cases. Layer 2 solutions \cite{gangwal2023survey} aim to work in a complementary way to the existing blockchain design. Such approaches include payment \cite{green2017bolt,avarikioti2018towards} and state channels \cite{negka2021blockchain}, rollups \cite{thibault2022blockchain}, sidechains \cite{back2014enabling, singh2020sidechain} and plasma \cite{poon2017plasma}. Those approaches have demands of a smaller scale and therefore are much easier to implement.

The present paper introduces Origami, a novel state channels design. State channels aim at reducing on-chain interactions for a given application. The main concept is to enable a given set of users that interact with an application deployed on a blockchain network to carry out most of this interaction off-chain among themselves and only make on-chain transactions when this is absolutely necessary. The main building block of such schemes is a mechanism that enables security guarantees equally strong to the ones offered by having all interactions happen on-chain. State channels maintain a state for the deployed app and each off-chain interaction produces a new instance of this state signed by all members. Each state progression can be validated on-chain, while inactivity of participants is also controlled by the mechanism.

State channels have been criticised in the past because a number of restrictions are set for an app to be state channels compatible. The set of participants has to be fixed, the order under which those participants interact with the app has to be strictly defined and a new channel is required for each new combination of participants in an application. Other solutions have gained momentum in the layer 2 ecosystem and mainly the roll-ups have been identified as the main mechanism that will enable public blockchains' scaling.

Given the volume of usage that wide adoption of blockchain technology will bring, it is sure that none of the current solutions is adequate enough. State channels do not interfere with the operation of the network but minimise the number of transactions requirements for a given application. Because of that state channels is a solution that can be easily combined with other approaches such as roll-ups and cumulatively expand the capacity of a given blockchain network.

Origami the state channels' design proposed in this paper aims to counter the limitations that have so far prevented state channels from being widely adopted as an effective scaling mechanism. Novelties introduced in Origami are:

\begin{itemize}
    \item Origami is the first state channel design to provide an extensive analysis of how unordered communication can be achieved in a state channel, and thus enable the off-chain execution of non turn-based apps.
    \item Origami is the first state channel design that allows flexibility in a channel's participant set. Once in the ecosystem, members can join and leave groups upon request without any on-chain interaction.
    \item Origami does not force groups and channels to close as a result of unresolved inactivity disputes. The channel adapts in order to progress upon the interaction of the rest of the members (excluding the misbehaving member).
    \item Origami offers a single point-of-entry into an ecosystem that will then allow participants to run any number of applications amongst any composition of Origami members. This ecosystem has the ability to infinitely scale, and therefore the ability to eventually support an environment of infinite state channel users that interact with each-other without the cost and burden of on-chain interactions. Origami is also exempt from restrictions placed by virtual channels, like an upper bound in funds that can be committed into a channel and the requirement that the participants must have a common intermediary.  
\end{itemize}

The rest of this paper is structured as follows: Section \ref{sec:related} provides information on other state channel designs of general purpose. Section \ref{sec:preliminary} presents the main background concepts for this work such as state channels and cryptographic accumulators. Section \ref{sec:design} outlines and analyses the components and functionalities of the Origami ecosystem.  Section \ref{sec:evaluation} compares Origami to existing state channel designs upon many different parameters. Section \ref{sec:security} presents the security guarantees of this design through the use of the UC framework. Finally, Section \ref{sec:conclusion} presents an overview of the results of this work and future work plans.

\section{Related Work}
\label{sec:related}

\subsection{State Channel Designs}

\textbf{Counterfactual: Generalized State Channels}

The authors of Counterfactual \cite{coleman2018counterfactual} have designed a generalised state channel template that can be utilised by any developer without requiring them to know how to interact with the blockchain. Through a simple API interested parties can program additional functionality to add on a channel which can then be utilised by any application that checks the requirements. In this way costs of deployment can be brought down by being split amongst app developers.

The main weak points of the framework are the ones shared by most state channel architectures: griefing and stale update posting. Measures can be taken against both but do not guarantee security against those behaviours, only limitation, often at cost of other functionality.

The whole concept is based around the term counterfactual. Outcomes , local and channel states and contract instantiations are implemented in this manner. Counterfactual X is the outcome that any participant can finalize but hasn't, while everyone can act like it has been realized on chain. Counterfactual state is defined as a state not yet brought on chain, but one that can get there by at least every participant it affects. Even the counterfactual instantiation of contracts is implemented, through a process of defining a contracts address and mapping the counterfactual variant to its ethereum smart contract one.
Counterfactual also implements Metachannels, the terminology being used to identify channels between two users formed through common intermediaries. Widespread third party functionalities like watchtowers have been taken into account and are compatible with the framework.

\textbf{ForceMove: an n-party state channel protocol}

The authors of this work \cite{close2018forcemove} identify three main reasons that can cause a dispute between participants in a state channel: Conflicting Moves, External States and Inactivity. Wanting to combat the last reason, they have designed their application in order to only concern turn-based applications, and store properties needed for the distribution of assets inside states themselves, effectively eliminating the first two causes of dispute. Their targets are mainly game or game-like applications. For Inactivity, they have implemented the force-move protocol. Its purpose is that a party, when faced with an unable or unwilling to cooperate opponent, can ensure their funds are not indefinitely trapped in the blockchain. When triggered, the force-move mechanism allows a time window for the challengee to respond in a number of possible, but predefined, ways. Any of those responses, or the challenged party's continued inactivity till the dispute period has passed, will result in either the continuation of the game, or its termination. In the last case, the funds will be split according to the last valid state. The force-move protocol requires an adjudicator that is usually, but not necessarily, an on-chain smart contract. This has to be taken into account by the application developers as they need to follow the supplied interface during development and provide a library that helps the adjudicator judge whether a state/transition/message is valid. Beyond monitoring the dispute process, the adjudicator also manages the assets of the game's participants. The major drawbacks of this protocol are defined by the authors. The design enables the spamming of the players with force-move challenges in various ways. While suggestions to prevent such incidents are provided, they do not fully secure the system against them. 

\textbf{Two-Party State Channels with Assertions}

In this work \cite{buckland2019two}, a construct for a basic form of state channels is presented. The authors create a design that aims to limit the cost for the on-chain interaction necessary to challenge an invalid state for honest parties. That is done by requiring a bond to be submitted by the party that asserts a state. This bond will be used as a refund if the counterparty proves this assertion to be invalid. A challenge can be issued in response to an invalid state, or at the face of inactivity, and is resolved by the submitting of an assertion signed by both parties. Additionally, authors attempt to make the size of the state uploaded on chain in dispute constant, by only demanding a hash of it to go in the contract. Apart from the contract of the corresponding application, the StateAssertion Contract (SC) also needs to be deployed on chain. The design concerns strictly turn-based applications between 2 parties that do not allow exceptions to be thrown by the application contract. In addition to those limitations, there is also a concern for privacy, as challenging requires the state to be submitted to the contract in plaintext. Parties also cannot benefit from services that free them from the need to be constantly online (eg watchtowers) as they would not be able to submit a state on their behalf. 

\textbf{Hydra: Fast Isomorphic State Channels}

The Hydra \cite{chakravarty2020hydra} design forfeits the sequential transaction processing that state channel designs usually function by in favour of concurrent processing enabled through the use of the Extended UTXO model. The isomorphic channels the authors describe, called heads, function by moving a set of UTXO's the participants decide on off chain, evolving them there and then making sure the latest state of the channel at close is reflected on the chain.The process begins with an initial transaction through which the parameters necessary for the channel (eg. participant list) are defined, and participation tokens necessary for the state validation process are forged for each participant. State propositions are validated through multisignatures, and periodically collected into snapshots to align the channel view of the head parties, that often differs due to the concurrent transaction confirmation scheme. Snapshot leaders are responsible for this gathering of states and for conflict resolution that may be needed. At the channel's closing, a contestation takes place during which channel members submit valid states more recent than any already submitted, and the latest one gets reflected on chain after finalisation of the head. This design results to isomorphic channels that are faster than any previously implemented ones and boast performance near physical limits.

\textbf{You Sank My Battleship! A Case Study to Evaluate State Channels as a Scaling Solution for Cryptocurrencies}

Kitsune, the framework presented in this work \cite{miller2020you}, is an application-agnostic, n-party state channel architecture that combines the features authors found most useful in previous implementations like Sprites and Perun. It provides a template for Application Contracts that facilitates the addition of state channel features into any existing application, and functions simply by freezing app functionality on chain and transferring it to the channel, and resumes it when the channel comes to a close. During the running of the app on the channel any party can propose an update and has to collect signatures from all other parties for it to be considered valid. The paper does not address how a scenario where updates are proposed simultaneously is handled. The channel can be closed cooperatively or through dispute. Disputes can be started by any participant and initiate a timer for parties to submit state information. Anyone can resolve the dispute, and it seems its only possible outcome is giving the state to the application contract and resuming on chain. 
Authors conduct an experiment during which they analyse the process, challenges, costs and usefulness of implementing a Battleship game through state channels, to reach the conclusion that any application that has a liveness requirement and not all participants are willing to cooperative is not particularly suited to this technology.

\subsection{State Channel Network Designs}

The environment around state channels has moved on from the goal of establishing channels between parties that require a deposit of funds on chain every time one is created. The aim is now to construct networks that will enable channels to be funded through each other, further restricting the load put on the blockchain and  increasing the use cases where state channels are applicable.

\textbf{PERUN: Virtual Payment Channels over Cryptographic Currencies}

Virtual channels, that are now a staple in state channel discussions, were first introduced by a concept in PERUN's \cite{dziembowski2017perun} virtual payment channels. The authors implement those channels in a recursive manner, standing on top of general state channels. They proceeded to design multichannels, that were targeted to the framework's needs, and essentially are state channels that have the ability to run many contract instances, called nanocontracts, in parallel, and hence support many payment channels concurrently. This feature has forced some more detailed mechanisms to be developed in order to make parallelism work. Authors have broadened general state channels with a conditional update mechanism that introduces greater flexibility concerning how and when the state will be altered. Additionally, greater care has been taken concerning the handling of participant stake funds to ensure that no over or double spending occurs in the multichannel.

\textbf{Sprites and State Channels: Payment Networks that Go Faster than Lightning}

Even though the main focus of the Sprites \cite{miller2019sprites} proposal is payment channels and networks, they base their design on a modular approach leaning on a general state channel construction. State channels are mainly used in the dispute process, the sequence of which has been adopted by most following state channel designs: a party raises a dispute after malicious behaviour occurs, a time period follows during which evidence is submitted, and finally, depending on the situation, the resolve is either cleared off chain or resolved on it.

\textbf{General State Channel Networks}

An extensive, formal definition for state channel networks, complete with security specifications had been missing until the authors of this work \cite{dziembowski2018general} provided it. A design that allows the recursive building of 2-party virtual channels, that do not require contact with the blockchain to open or close, across any number of intermediaries, that still offer the same functionality and guarantees as a ledger channel is what this work contributes. The authors describe the creation and optimistic update of channels in a constant number of rounds, while the optimistic execution is relative to the span of the channel.

Channels are formed after the intermediary is contacted and has accepted each party's request. The safety of the intermediary's locked funds is guaranteed by contract instances run on the underlying ledger channel with each of the participants. Those instances are also used in the case of a dispute, which is handled by the non-trusted intermediary before resorting to the blockchain, thereby adding an extra layer of defence.

Although the balance of the intermediary is completely secure, there is a number of requirements, the burden of which the authors suggest be mitigated by a service fee.

\textbf{Multi-Party Virtual State Channels}

The authors of this work \cite{dziembowski2019multi} offered two significant contributions to state channels existing so far at the time of publication. 

Introduced in this work are multi-party virtual state channels, an expansion on 2-party virtual channels that had already been implemented. Virtual channels can be opened and closed without ever referring to the blockchain. Multi-party virtual state channels can execute contracts that concern over 2 parties. They stand and are built recursively on 2-party ledger channel networks, through which all participants must be connected, no creation of channels of intrinsic greater length is supported. For every multi-party virtual channel created, participants must install in every sub-channel contract instances that guarantee that the intermediary's funds are safe and that the outcome of the virtual channel will be updated on said sub-channels.

The second major contribution is the introduction of direct dispute state channels. Those introduce a functionality that redesigns the dispute process so that participants refer to the ledger as soon as possible after an honest party identifies possibly malicious behaviour, instead of contacting the intermediary. The dispute board is an on-chain component where states are uploaded to in case of dispute to be compared and the valid one is finalized. This approach makes the worst case time complexity of the channel independent of its length, and allows dispute outcomes to be seen and used by other contracts. However, it can cause serious transaction load on the blockchain depending on the scenario, since all parties can be required to upload their state view for a dispute. Channels that implement direct and indirect dispute processes can cooperate seamlessly in whatever proportions serve each application best.

\textbf{Nitro Protocol}

 In the Nitro protocol \cite{close2019nitro}  design, it is possible to construct virtual state channels that function following the same rules as directly funded channels do, and on top of that the state of a channel can change from virtual to direct at any point. All channels of the network function independently, a crucial prerequisite for the core reasoning process of the protocol regarding the validity of updates.  The authors identify the key parts in the extraction of value from a channel, and go on to split this process in two phases. Finalisation, which is the storing of an outcome on the chain, and Redistribution, the updating of balances to reflect said outcome. In Nitro, this equals updating the single contract that manages all balances per network, the Adjudicator. Nitro relies on the reasoning that if an outcome can be reached, then that knowledge is sufficient to support indirect funding in channels without the actual execution of on chain operations that finalise it being necessary. Indirect funding can happen in two ways. Ledger channels exist to provide funding of sub channels and participant accounts, while virtual channels can be created in Nitro through a common intermediary between participants.

\subsection{Connectivity Solutions}

\textbf{Pisa: Arbitration Outsourcing for State Channels}

PISA \cite{mccorry2019pisa} builds upon the functionality of Monitor and Watchtowers proposals, with the same end goal, eliminating the requirement for a state channel participants to be constantly online, or be left vulnerable to unfortunately timed disputes and execution fork attacks. PISAs contribution is the option for a participant to employ a third party to watch over the channel in their absence, while providing the participant with evidence, in the form of a cryptographic receipt, that will act as a fail safe if the hired third party does not act as intended. The process is fairly straightforward, with a potential custodian setting up a contract in which they deposit a large sum to act as collateral. Customers pay for services through one-time payment channels and receive a receipt, and the custodian either settles any occurring disputes or has their deposit burned. The protocol ensures the privacy of the state by only providing the custodian with a salted hash of it, and ensures fair exchange through the nature of payment channels and giving the custodian the tools to verify the validity of incoming conditional transfers. However, in the case of a custodian being offered a payout larger than their stake, they cannot be deterred from colluding with the rest of the parties in the absence of the customer. An effort is made by the authors to mitigate this fact by reserving a part of the custodians collateral to compensate the cheated customer, but this gives the custodian a way to limit their loses and hence have increased motive to misbehave.

\textbf{BRICK: Asynchronous State Channels}

Brick \cite{avarikioti2019brick} addresses the fact that most current implementations require a synchronous network to be able to provide safety in the channel environment and that surrenders to malicious parties valuable information, like the exact time the network needs to be controlled by an attack, a problem that is often undermined as many proposals evaluate their frameworks while assuming perfect substrate on the blockchain that does not correspond to real world circumstances. 
Brick claims to be the first state channel architecture that can maintain security in a fully asynchronous network by redesigning the dispute process to take place off chain, thus lifting the constrictions imposed by congestion on the chain that could interfere  with the resolve process and the need for members to remain constantly online or risk losing their assets. A key element of this framework is the committee. Introduced to function as a validation mechanism, its members maintain power of attorney over the channel and receive a broadcast of the state after every update.  Instead of starting disputes, the members can either cooperate or reach out to the committee to move the process along. Committee members are sufficiently incentivised through rewards and locked collateral to behave honestly and ensure proper functionality in the channel when called upon. BRICK succeeds in guaranteeing the security, privacy and liveness in a state channel without the use of time windows and disputes and with a satisfactory overhead in terms of time penalties, but does increase the cost for channel participants as they have to provide the committee member rewards.

\section{Preliminary Concepts}
\label{sec:preliminary}

\subsection{State Channels}

A major factor that limits the scalability of most popular blockchain networks is the requirement for every node to execute every transaction. Even though this has been deemed necessary for security reasons, it is logically redundant to associate all network nodes with transactions that are not relevant to them. 

Payment channels were the first approach to be based on this premise, aiming to transfer interactions off-chain and execute them strictly between the directly involved nodes, while maintaining the same security guarantees as on-chain transactions on a blockchain network. Payment channels implemented this approach strictly for transactions related to payments and cryptocurrency transfers.

State channels were an extension of this idea, beyond payments and into all state transitions, allowing smart contracts to be executed in this manner. Transferring communications off-chain and confining them between the directly interested users offsets the cost and time overhead that would occur were these interactions to take place on-chain.

On-chain interactions with a smart contract depend on the consensus mechanism to guarantee that any given state transition's timestamp, content and validity is public and indisputable. A state channel operates by transferring the execution of the contract to a local level, just between the participating users. Total consensus is needed between said participants for the state update to have the same validity as one issued on-chain. Additionally, to maintain security, it must be assumed that every channel transaction can be published on-chain at any point. Therefore, state channels greatly depend on the availability of their participants and the liveness of the blockchain system.

A major component of a state channel design is the channel management contract, a smart contract instantiated on-chain to define the functionality of the state channel through public, immutable code. In said smart contract, users deposit the funds they wish to use within the channel as well as an additional amount to serve as a stake. In the pessimistic case where channel participants diverge from acceptable behaviours, the contract can act as an arbitrator and if necessary withhold part of that stake. Therefore, a state channel does not have to assume the lack of malicious or unavailable participants.

In the optimistic case, participants communicate with each-other to progress from one state to the next and only interact with the blockchain network to deposit and withdraw assets. 

The phases in which a state channel may be in during its operation are the following:

\textbf{Opening Phase:} Often also called the Funding phase, this is the protocol followed for the creation of the state channel. During this phase, participants deposit their stake and any assets they wish to use within the channel. Participants also determine and sign the starting state for the state channel.

\textbf{Update Phase:} The core functionality of a state channel takes place during the Update phase, that allows moving on from a given state to the next. In a typical update round, through cryptographically signed messages, state update proposals are broadcast from one participant to the rest. A valid state gets signed by all channel members and therefore has the same finality as a state transition happening on-chain. 

\textbf{Dispute Phase:} The dispute phase is the regulation mechanism of the channel against behaviours that diverge from the protocol. Since full consensus is required for the update process to proceed, unresponsive behaviour is especially disruptive to the channel's functionality. Similarly, invalid state proposals are not acceptable and are treated as inactivity. In such cases, participants rely on the channel management smart contract that is deployed on the blockchain network to resolve the situation. Different state channel designs show the most diversity in the way they handle disputes. Depending on the outcome of this phase, a state channel may go back to the update phase or be forced into the closing phase.

\textbf{Closing Phase:} Dissolving a state channel is a process that follows a particular protocol. It can be triggered either by an unresolved dispute or by participants agreeing upon the termination of the channel. In both cases, a valid closing state must be passed on to the management smart contract that will proceed to divide the stored assets between the participants according to said closing state.

Properties that cannot be missing from a functional state channel are those of finality and trustlessness. It must be ensured that the safety of any participant's assets does not depend on the compliance of the other channel members. It must also be guaranteed that any state reached within the state channel following proper procedure has equal weight to one reached on-chain.

Through the described mechanisms and several variations of it, state channels manage to decongest the blockchain network by enabling strictly off-chain communication in the optimistic case, and reduced on-chain transactions in the case of malicious behaviour.

The majority of existing state channel designs set a number of specifications that must be met in order for an app to be state channel compatible. A number of them seem to be hindering the the adoption of state channels in practical scenarios. Said requirements are:

\begin{itemize}
    \item Constant Connectivity: Derived from the aforementioned dependency on the availability of participants, state channels demand all users to be responsive at all times and categorize unavailability as malicious behaviour. However, it is not a realistic demand as it is not practically feasible to keep track of a state channel at all times, or to be able to control outside factors that might prevent responding.
    \item Countdowns: In order to set a window after which a participant is considered unresponsive, timers are frequently used in state channel designs. Unfortunately, time measuring in blockchain networks has yet to be done in a way that is both secure and highly accurate.
    \item Ordering: Still related to the participant availability issue, most state channel designs force a predefined order of communication and therefore exclude any possible application that does not function under such a premise.
    \item Static Participant Sets: The user set of the state channel is usually the one defined in the funding phase and it remains unchanged. This lack of flexibility results to the necessity for a new channel to be created for every, even slightly, diverging group of users. It also prohibits applications that function with a changing set of participants from making use of state channels.
\end{itemize}

\subsection{Accumulators}

Accumulators are compact data structures that include a set of elements and can provide their inclusion proofs without revealing them.  For every value of the set a witness proof can be produced and thereupon determine whether the value is incorporated in the accumulator or not.\cite{kumar2014} Used as a cryptographic commitment scheme for a set of elements $ S= {(e_1, e_2,,..e_n)}$, an accumulator is a useful tool that enables the prover, who stores the entire set, to convince any verifier, who only stores a succinct digest of the set, of various set relations. 
 
Accumulators improve efficiency in terms of used storage and time, even when handling an arbitrarily large set of values since the size of the data structure remains constant. RSA accumulators specifically, offer the advantage of constant-size proofs. Numerous designs have been proposed combining a variety of features an accumulator can present, such as the size of the initial set, the type of membership proof, the existence of a trusted coordinator, known as accumulator manager, and the required update frequency for the participants \cite{ozcelik2021overview}. 

In this section, we present some of the main characteristics and cryptographic assumptions of an accumulator scheme, that highlight its utility in the proposed protocol where it is used as a commitment scheme for the openings of sub-protocols. \ref{subsec:basechannel}   

Trapdoors in cryptography consist of information needed to perform the inverse cryptographic operation\cite{ozcelik2021overview} coordinated by the accumulator manager. In recent works \cite{Boneh}, \cite{Lipmaa2012SecureAF}, \cite{Wesolowski}, \cite{Dobson} efficient schemes have been proposed without requiring a trusted setup. 

Following our previous work \cite{NEGKA2023100114} we built upon the assumption that the RSA generator has been generated through a secure function based on Wesolowski’s \cite{Wesolowski} Adaptive Root Assumption and the related work of Boneh et al. \cite{Boneh} that showed that the Wesolowski’s proof is a succinct proof of knowledge of a discrete-log in a group of unknown order. 

\paragraph{Strong RSA Assumption} The Strong RSA Assumption states that given a random generator of unknown order $g \in G$, it is infeasible to find any root of it i.e. an integer $l \in Z$ and an element $u \in G$ such that $g^{1/l}= u$. Strong RSA assumption generalizes and implies the RSA assumption and constitutes the core of numerous cryptographic procedures. 

\paragraph{Adaptive Root Assumption} Wesolowski in \cite{Wesolowski} states that it is difficult to find a random  root of a chosen group element. Those two assumptions are incomparable, as the latter states the difficulty of finding a random root of a chosen group element, whereas the former upholds the difficulty of finding a chosen root of a given random group element \cite{Boneh}.

\paragraph{Accumulator's functions} This approach includes the existence of a trusted manager and users responsible for their elements and the corresponding membership witness acting as provers or verifiers at a specific verification round. \cite{Baldimtsi2017}
\begin{itemize} 
 \item \textbf{Generation/Set up Algorithm} generates the initial set of the accumulator \(A_0\). In the proposed protocol the initial accumulator is generated with a generator $g \in \mathbb{G}$ where the Strong RSA assumption holds. Let $H_{prime}$ a hash function that maps any element $x_i$ to a unique odd prime number $e_i$ and $g$ the generator of every element $x \in \mathsf{S}$ , $\mathsf{S}= \{x_1, x_2,...,x_{(n)}\}$ and $e_i=H_{prime}(x_i)$. The accumulator of \(\mathsf{S}\) of the initial elements after being mapped with the hash function $H_{prime}$ is constructed as:
    
    \begin{equation}
        A= acc(S) = g^{\Pi_{i\in [n]}e_i}
    \end{equation}
 
 \item \textbf{Add Algorithm} produces the updated accumulator value $A_{t+1}$, after the addition of an element $x$ to the accumulator, and the membership witness proof for element $x$, labeled as $w_x$. Additionally, an update message is produced $upmsg_{(t+1)}$ that enables the accumulator's users and proof holders to update the witnesses of their elements. 
    \begin{equation}
        A_{t+1}= (A_{t})^{e_i}
    \end{equation}

 \item \textbf{Del Algorithm} respectively produces the updated accumulator value $A_{t+1}$, after the deletion of an element $x$ from the accumulator, and the non-membership witness proof for element $x$, labeled as $u_x$. Accordingly, an update message $upmsg_{(t+1)}$ is produced to inform the accumulator's users in order to update the witnesses of their elements.
In the proposed protocol, we assume that the owner of an accumulated element $e_i$ maintains the corresponding membership witness $w_i$, which equals the value of the accumulator before the aggregation of the element, the process of updating the accumulator is presented in Equation \ref{eq:del_el}.
    
    \begin{equation}
    \label{eq:del_el}
        A_{t+1}= {w_i}_t
    \end{equation}
    
 \item \textbf{Create Membership/Non membership Witness Algorithm} produces respectively the inclusion or exclusion proof for an element. Membership witness $w_j$ for an element $e_j$ is computed with:
    \begin{equation}
        w_j= (A_t)^{-e_j}
    \end{equation}
    
    As presented in \cite{NEGKA2023100114} to integrate accumulators into state representation for state channels, the user that adds an element to an accumulator is burdened with the responsibility to maintain its previous value $A_{t-1}$ as the exponentiation of the accumulator by $e_j^{-1}$ cannot be efficiently executed in a hidden-order group. As for the non-membership witness, $u_i$ of an element $e_i$, that is not included in an accumulator $A$ can be calculated using the Bezout coefficients when the product of all the accumulated elements is known.

 \item \textbf{Update Membership/Non membership Witness Algorithm} updates the membership/Non membership  witness for an element $x$ after the addition or deletion of an element $y$ to the accumulator.Updating membership witnesses of an element $e_j$ upon the addition of an element $e_i$ is given simply by adding the element $e_i$ to the witness proof ${w_j}_t$.
    \[  {w_i}_{t+1} = {w_i}_{t}^{e_i} = {A}_{t}^{e_i}\]
    
Whereas updating membership witnesses for $e_i$  after the removal of an element  $e_j$ requires the computation of $e_j$th root of $A_t$ which corresponds to the updated witness. After the computation of the Bezout coefficients $a,b$ we can produce the updated membership witnesses according to the following equation: 
    \begin{equation}
        {w_i}_{t+1} = {w_i}_{t}^{b}{A}_{t+1}^{a}
    \end{equation}
 
 \item \textbf{Verification Algorithm} is executed by any user that preserves the latest state of the accumulator against which the existence of an element $x$ in the accumulator can be verified using its membership witness $w_x$. For the verification of a membership proof $w_i$ of an element $e_i$ given the current accumulator state $A_t$ one exponentiation in $\mathbb{G}$ is required. Adding the element $e_i$ to the set accumulated in $w_i$ and checking if the result equals  $A_t$, verifies the proof.
    
    \begin{equation}
        A_{t}= (w_i)^{e_i}
    \end{equation}

    Respectively a non membership witness $u_i=( a, g^b)$ for an element $e_i$ can also be verified as in the following equation: 
    \begin{equation}
    \label{eq:ver_non_mem}
        A^a *( g^b)^{e_i} = g
    \end{equation}

\end{itemize}

\section{Design}
\label{sec:design}

\subsection{Concept}

\begin{figure}
    \centering
    \includegraphics[width=0.5\linewidth]{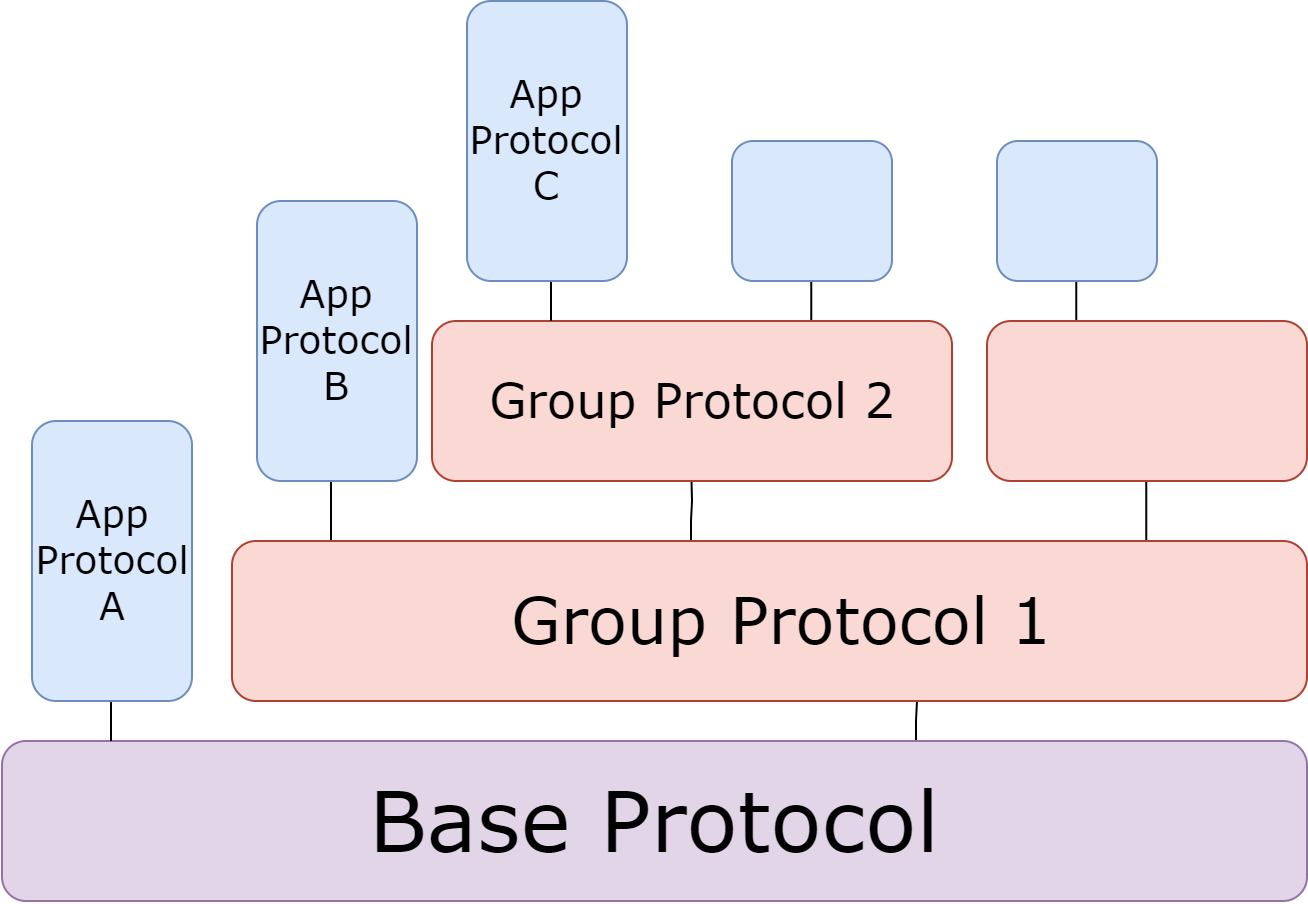}
    \caption{Origami concept.}
    \label{fig:origami_concept}
\end{figure}

Origami design is based on the concept of building off-chain interaction protocols (state channels) upon other similar protocols, in order to allow the creation of constructions characterized by the required level of flexibility. There are three types of off-chain interaction protocols, multiple instances of which are used to construct the origami state channels environment. In Figure \ref{fig:origami_concept} an abstract schematic view of the Origami environment is depicted. Each one of the blocks represents an instance of an off-chain interaction protocol which (apart from the case of the base protocol) is based upon another protocol represented by the block directly underneath it. The base protocol (single instance) is similar to a traditional state channel and enables users to deposit funds through on-chain transactions and join the Origami environment. The group protocol is similar to the base protocol with the main difference that it has to be funded on either the single base protocol instance or on another instance of the group protocol. The main functionality offered by the group protocol instances is that they enable participants to form smaller and more flexible subsets of users according to their needs, without having to do any on-chain interaction. Finally the app protocol instances are based upon the single base protocol instance or an instance of the group protocol and are application specific thus enable users to run applications off-chain. 

The Origami approach facilitates upon this hierarchical construction of multiple off-chain interaction protocols to support multiple features that are missing from existing schemes, while keeping the required on-chain interaction of the users at a bare minimum. The name of the design (Origami) comes from the Japanese art of creating paper structures through multiple foldings of a piece of paper, a process to which our multiple off-chain protocols resembles. The intuition is that a large number of users can join the Origami system, by taking part into the single base protocol instance. As subsets of those users want to interact, they will tend to form instances of group protocol and eventually instances of the app protocol, for which the set of participants is going to be driven by the requirement for interaction. The lesser the participants in a group/app instance are, the more efficient that gets in terms of operation (e.g. number of required signatures to progress state, probability of inactivity or bad behavior etc.) The state of both base and group instances is strictly defined. It relates to a balance-sheet with the funds of each participant in the specific protocol and binding information about the protocol instances on top of that to which part of the funds have been committed. On the other hand the state of the app instances is application specific and is defined in ad-hoc way according to the application that is going to be deployed on top of such an instance.

The proposed design aims to support an ecosystem of users that only go through the process of joining the system once (through one on-chain transaction), but can then run any number of state channel compatible applications with any subset of the rest of the participants (with no on-chain interaction in the optimistic case). 
Central to the architecture are the following protocols:

\begin{itemize}
    \item \textbf{The base protocol} is used as the initial step that allows users to join the ecosystem and deposit funds.
    \item  \textbf{The group protocol} is a scaling mechanism that aims to decrease any overhead that is bound to occur as the ecosystem grows. 
    \item  \textbf{The app protocol} through which a subset of members of the ecosystem can operate an independent channel, created with the intent of running a state channel application.
\end{itemize} 

Apart from the base protocol that is directly attached to an on-chain smart contract and requires users to deposit funds in it, the rest of the protocols operate completely off-chain for the optimistic case and with minimal on-chain interaction in the pessimistic case. 
The functionality of each protocol along with the interactions between their instances are extensively analysed in the present Section. The terms channel and protocol are used interchangeably. 

In the subsequent subsections the three different protocols are presented. As the protocols are quite similar and their differences are restricted to specific parts of the their functionality, the base protocol is described in detail while subsequently we present the group protocol on the basis of its differences to the base protocol and the app protocol on the basis of its differences to the group protocol.

\subsection{Base Protocol}
\label{subsec:basechannel}

An important building block for Origami is the base protocol, an instance of which is identified as the base channel and operates similarly to the conventional notion of a state channel. It functions as the entryway for those who wish to join the ecosystem. All users that want to participate in Origami must become members of the base channel after depositing funds into its corresponding smart contract. 

Every phase and process of the base channel will be presented in this subsection.

\subsubsection{State representation}
\label{subssubsec:bcstate}

\begin{figure}
    \centering
    \includegraphics[width=\linewidth]{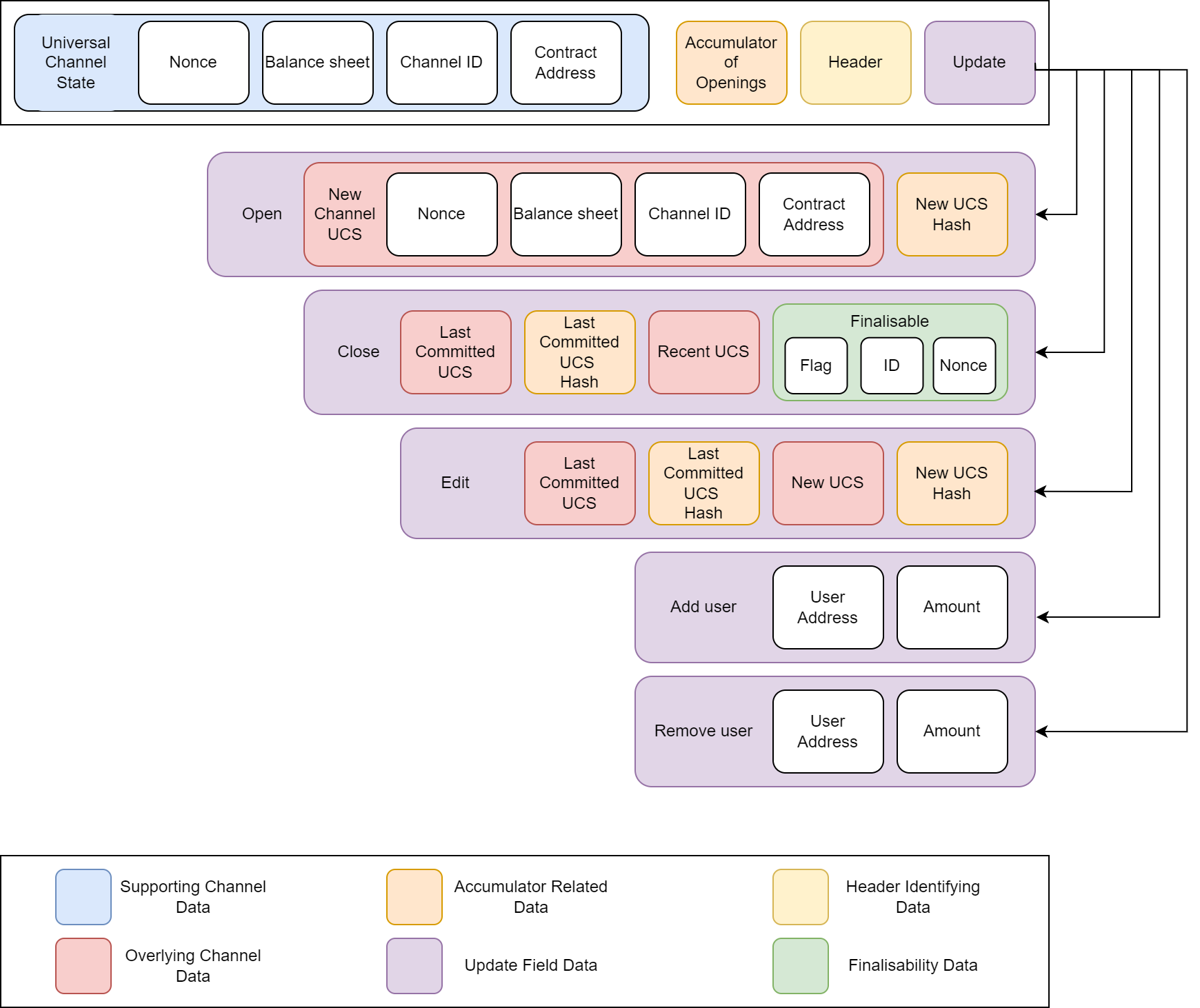}
    \caption{Base channel state structure}
    \label{fig:bc-states}
\end{figure}

Base channel state updates pertain to management related actions in the ecosystem, such as the addition and removal of participants, and the supporting of the operation of instances of the higher levels protocols, such as group and app channels. The state of the base channel is updated whenever there is change in the participants set (a user is added/removed from the ecosystem) or there is a change regarding the group/app channels that are deployed directly upon the base channel (a channel is opened, closed or changed). Therefore, the form of the state of a base state channel adheres to the following template, also seen in Figure \ref{fig:bc-states}:

The state of a base channel consists of three static fields and one dynamic field the form of which is dependant on the nature of the last state update. These fields are:

\begin{itemize}
     \item \textbf{Universal Channel State:} The Universal Channel State (UCS) is a set of channel-defining data that facilitates the transfer of information between protocols in different layers. It is separately signed before it is included in the state and contains the following information:
     \begin{itemize}
          \item \textbf{Nonce:} A counter that is used to order states in a channel and is useful when trying to prove a state is stale (a more recent valid state exists).
          \item \textbf{Balance sheet:} A mapping of the members of the channel to the amount of available funds they posses in the channel. Balance sheet apart from defining available funds, is also used as point of reference for the members list of a channel. 
          \item \textbf{Channel ID:} Each channel is assigned a unique identifier for management purposes. For the base channel (which is only one across the ecosystem) the channel id is specifically set to 0.
          \item \textbf{Contract Address:} Each channel, irrespective of its type (base, group or app) is based upon a deployed smart contract, that holds all the required functionality to support the channel (e.g. validation of state transitions). When a new channel is instantiated the address of the corresponding smart contract is specified in this field.
     \end{itemize}
     \item \textbf{Accumulator of Openings:} A structure that stores commitments to all open channels on top of the state base channel. For efficiency reasons and in order to keep the size of the state representation fixed the Openings structure is implemented as an RSA cryptographic accumulator. For every new channel that is opened upon the base channel an element (new channel's UCS) is added to the accumulator. If such a channel is closed then the corresponding element is removed from the accumulator. The new channel's participants are responsible for storing the element on their side and also for updating the corresponding membership proof of the accumulator for that element.
     \item \textbf{Header:} Origami follows a unique communication design that requires setting a member as a Header for each round. This enables the unordered state updates from channel participants. A state indicates who will be acting as the Header for the next state update round. The relevant process is analysed in Section \ref{subsubsec:comms}.
     \item \textbf{Update:} This is a complex field that indicates the type of state update being proposed and also holds all the required information for that update. There are five update instances which are either related to actions for channels based on top of the base channel: (a) opening a new channel (b) closing an existing channel (c) editing an existing channel, or related to user management of the base channel: (d) adding a user or (e) removing a user.

According to the type of the update the update field is structured as follows.

\begin{itemize}

\item For opening a new channel the sub-fields of the update field are :

\begin{itemize}
     \item \textbf{New Channel's Universal Channel State:} The Universal Channel State (UCS) is a set of channel-defining data that facilitates the transferring of information between protocols in different layers. It contains the following information:
    \item \textbf{New Channel UCS Hash:} a hash of all the fields in the UCS to form an element that can be added to the Accumulator and facilitate verification of the state transition.
\end{itemize}
 
\item For closing an existing channel the sub-fields of the update field are :

\begin{itemize}
    \item \textbf{Last committed UCS:} The last Universal Channel State (UCS) that has been committed (through an open or an edit state update) for the channel to be closed in the Accumulator of Openings.
    \item \textbf{Recent UCS:} The UCS that corresponds to the last state of the channel that will be closed.
    \item \textbf{Hash of last committed UCS:} The hash of the last committed UCS which is needed to proceed with the removal of the element from the accumulator. Including the hash of the UCS in the update field reduces the volume of on-chain operations required.
    \item \textbf{Finalisability tuple:} A tuple of three values that is used to communicate whether a channel can be closed at its current state to the supporting channel it is based upon. The three elements are:
        \begin{itemize}
            \item \textbf{Finalisable Flag} is a value set to true only for states deemed as appropriate to close the channel on.
            \item \textbf{Nonce} must be identical to the nonce of the most recent state the channel wants to close on.
            \item \textbf{Channel ID} must correspond to the ID of the channel to be closed.
        \end{itemize}
\end{itemize}

\item For editing an overlying channel (removing or adding a user) the sub-fields of the update field are :

\begin{itemize}
     \item \textbf{Last committed UCS:} The last Universal Channel State (UCS) that has been committed (through an open or an edit state update) for the channel to be edited in the Accumulator of Openings.
    \item \textbf{New UCS:} The UCS that corresponds to the updated data of the edited channel.
\end{itemize}
  
\item For adding a new user, given that the user has deposited an amount in the smart contract of the base channel, sub-fields of the update field are :

\begin{itemize}
    \item \textbf{User address:} The address of the user to be added.
    \item \textbf{Amount:} The funds the user has deposited and are being added for them in the channel's balance sheet.
\end{itemize}

\item For removing an existing user, given that the user holds a specific amount of the funds in the channel, the sub-fields of the update field are :

\begin{itemize}
    \item \textbf{User address:} The address of the user to be removed.
    \item \textbf{Amount:} The funds the user has and will be withdrawn from the channel.
\end{itemize}

\end{itemize}

\end{itemize}

\subsubsection{Communication model and disputes}
\label{subsubsec:comms}

Existing state channel designs assume that the ordering according to which the participants may propose state updates is strictly predefined. This mainly happens because having a single valid state proposer at each round significantly simplifies the process of validating the state proposals.

\begin{figure}
    \centering
    \includegraphics[width=\linewidth]{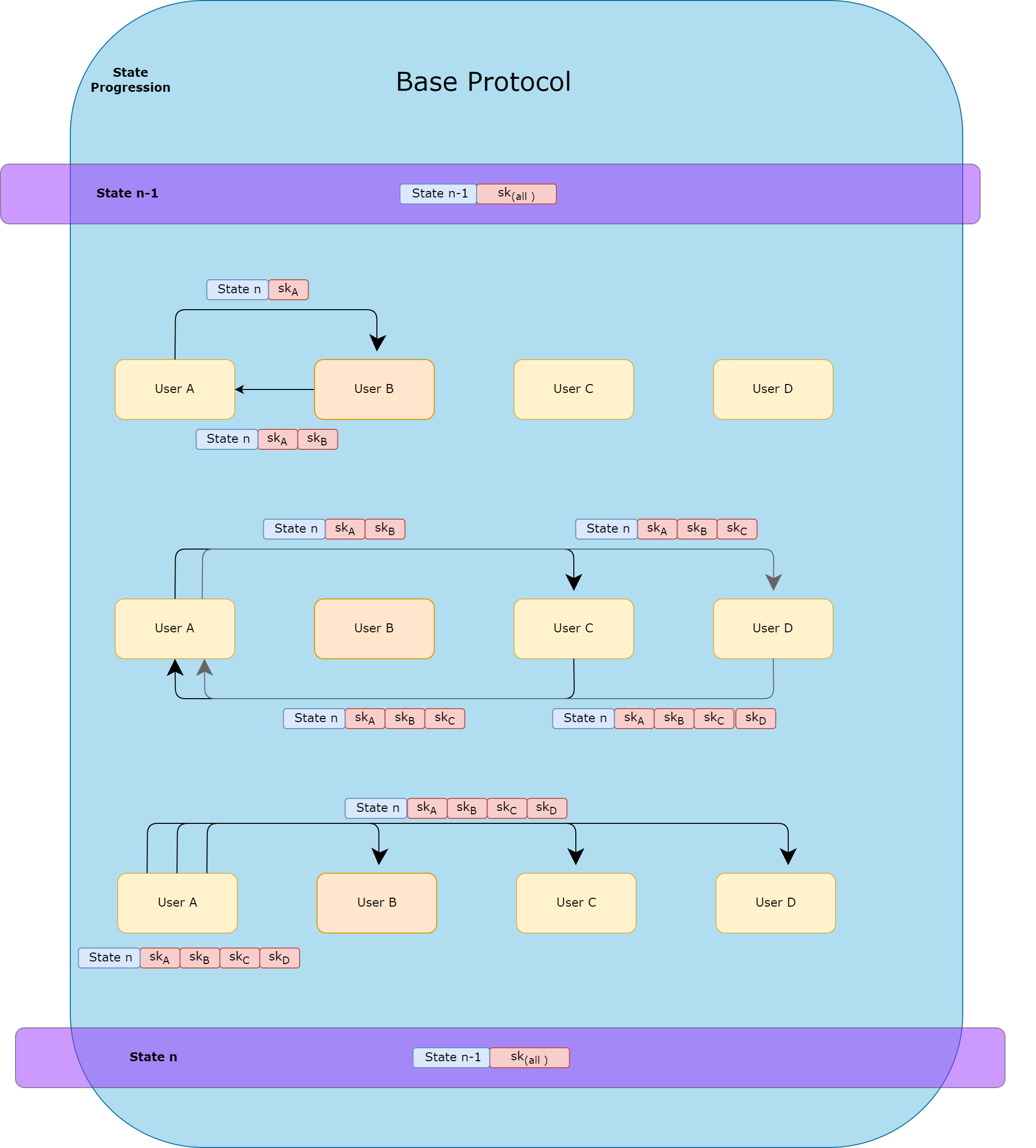}
    \caption{Depiction of the off-chain communication between participants.}
    \label{fig:communication}
\end{figure}

Because Origami design aims to eliminate such ordering restrictions in terms of which participant is entitled to make the next update proposal and lift the universal restriction of running strictly turn-based applications within the state channel, we have adopted a novel way to regulate in-channel communication. The process can be observed in Figure \ref{fig:communication}, and the analysis is based on the notations of the same Figure.

Lets define a round as the duration it takes for a new state to be established, ergo the process of going from state $n-1$ to state $n$. For each round, a channel participant acts as a header, in the sense that an update proposal must first pass through them. The header is determined through the order of the member entries in the balance sheet of the respective protocol, starting from the first entry and moving along by one each round. This ensures an equal distribution of this position among the participants.

The process of establishing a new state can be described in the following steps:

\textbf{Step 1:} Assuming User B is acting as the header, and User A intends to send a state update proposal, then User A must sign this proposal and first send it to User B. If this state update is valid, the header signs it and sends it back to the proposer.

\textbf{Step 2:} User A now broadcasts the proposal (signed by the header) to the rest of the participants. They, in turn, sign it and send it back to User A, only if it has already been signed by the round header. 

\textbf{Step 3:} User A, having collected all necessary signatures on their proposal, broadcasts it as the new state.

This method ensures that, while there is no sole participant able to suggest the next update, there will also be no confusion or conflict if various participants concurrently propose their own state update. In case the header receives more than one valid state update proposal in one round, then they are entitled to choose which one to progress with.

\begin{figure}
    \centering
    \includegraphics[width=\linewidth]{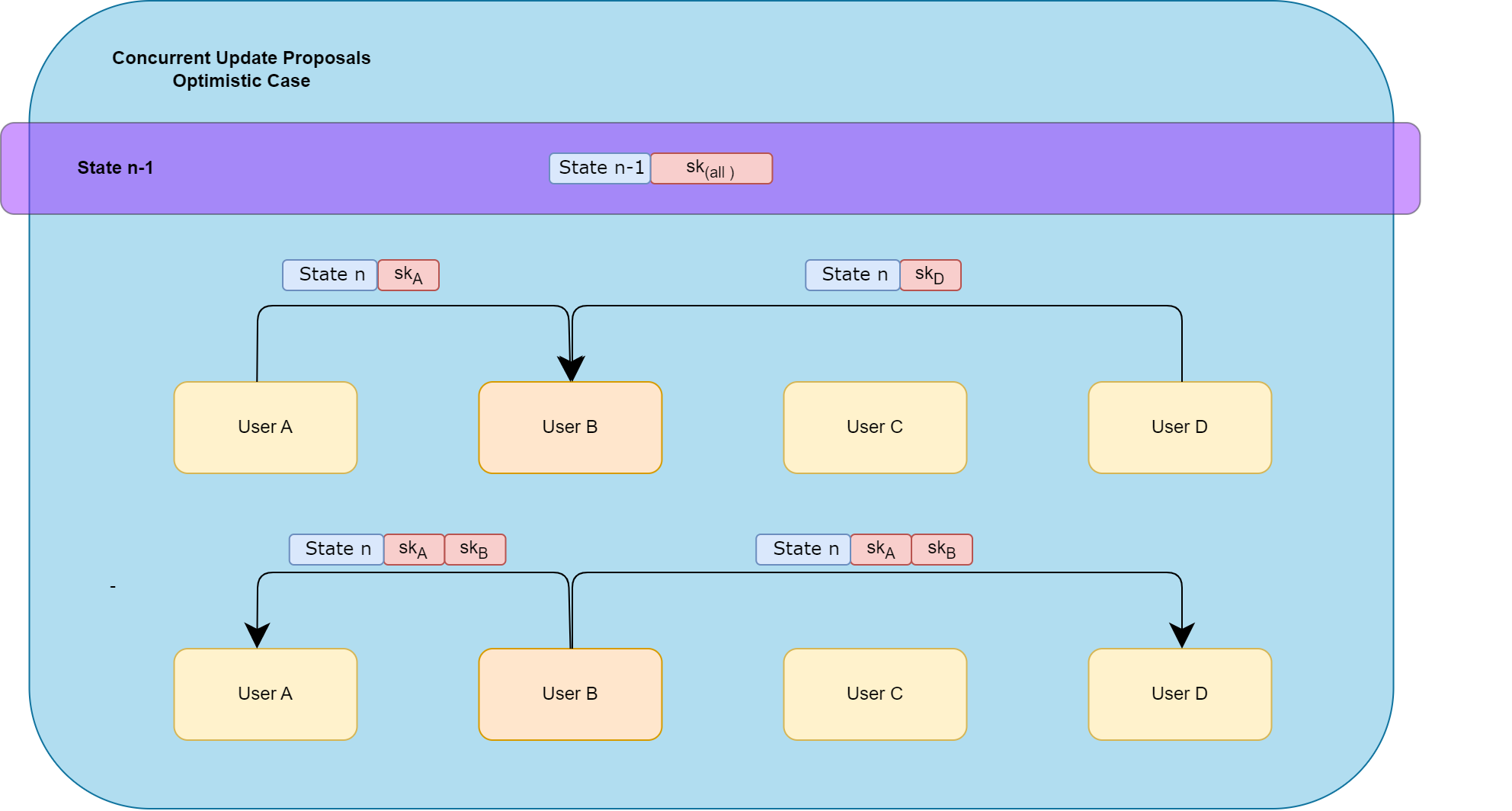}
    \caption{Optimistic case protocol for concurrent update proposals.}
    \label{fig:concopt}
\end{figure}

It is important for the rest of the channel participants to be able to distinguish between the header opting for another proposal and the header being inactive. Therefore, as shown in Figure \ref{fig:concopt}, the header has to provide all proposers with the chosen proposal, so that all involved parties know the channel is progressing even with an alternative state. 

In order to support the described protocol Origami includes two types of disputes, the header inactivity dispute and the channel member inactivity dispute.

\begin{figure}
    \centering
    \includegraphics[width=\linewidth]{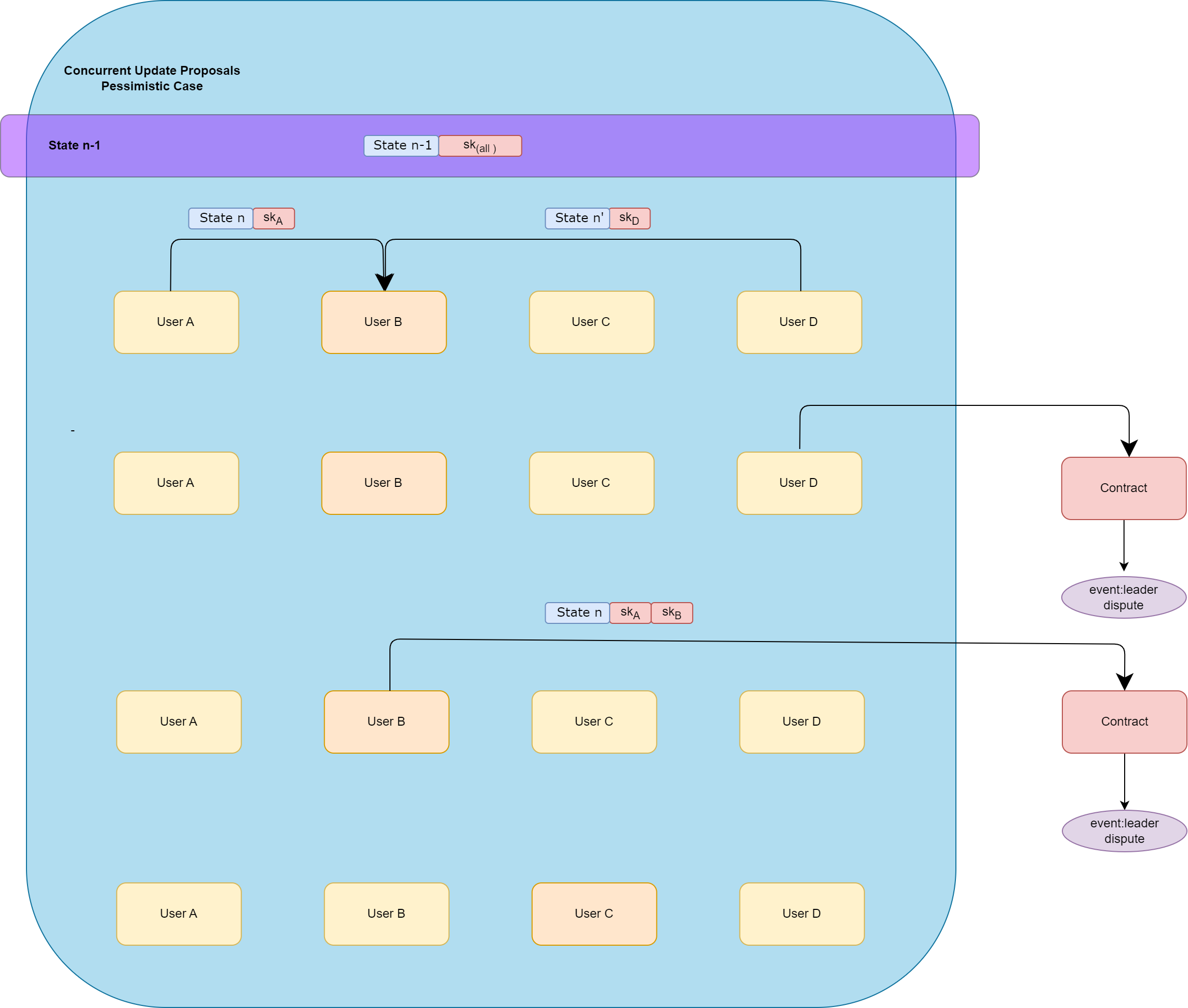}
    \caption{Pessimistic case protocol for concurrent update proposals.}
    \label{fig:concpess}
\end{figure}

\textbf{Header Inactivity Dispute}

If a state proposer does not receive a reply from the round header (either his or another participant's state proposal signed by the header), then they are entitled to go on chain, and initiate a header inactivity dispute, as depicted in Figure \ref{fig:concpess}. The process is as follows:

\textbf{Step 1:} A channel participant, in this case User D, sends a proposed state update to the round's header and receives no reply. User D is therefore not aware that User A has also sent a proposal, and they are entitled to start a dispute through the on-chain contract. 

\textbf{Step 2:} This dispute initiates a first countdown during which the header must provide a signed proposal they are proceeding with, whether that is the one from User A or User D. If this happens, then a second countdown is initiated to allow this proposal to be challenged in case it is stale. 

\textbf{Step 3:} If the group header provides to the contract a valid response to the dispute, then the successful proposer can retrieve their signed proposal from the contract and broadcast it to the other channel members and continue with the progression of the channel. Alternatively, if the group header never provides a valid reply to resolve the dispute, then they suffer a penalty and the contract moves the channel on to the next header.

\textbf{Channel Member Inactivity Dispute}

\begin{figure}
    \centering
    \includegraphics[width=\linewidth]{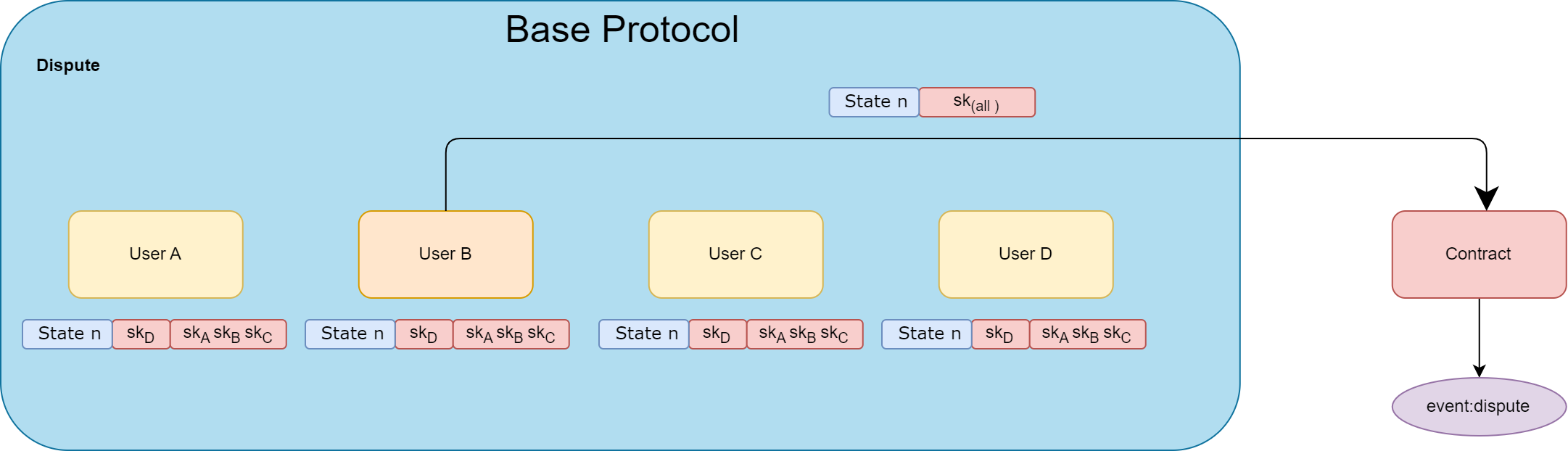}
    \caption{Dispute process against inactive participant.}
    \label{fig:bcdispute}
\end{figure}

If a member exhibits inactive behaviour by refusing to sign a valid state, a dispute against them can be triggered by any participant. The process is shown in Figure \ref{fig:bcdispute} and described by the following steps.

\textbf{Step 1:} The member most likely to start this process is the one that made the state update proposal. Said member therefore provides the contract with their proposal, signed by the header and by as many members responded, as well as the last valid state. The contract allows a window for either of those submissions to be challenged as stale.

\textbf{Step 2:} After deeming the state proposal is valid, the contract starts a timeout allowing a window for the inactive member to respond with a signed version. If this happens, then the dispute is resolved. If it does not, the channel removes the absent party from the process by no longer requiring their signature to deem a state as valid. The inactive party also suffers a penalty. This process is described extensively in Section \ref{subsubsec:bgremove}.

Part of resolving an inactivity dispute is judging whether or not the non-signing member is justified in their inactivity, which comes down to whether the proposed update is a valid transition from the previous state. This is determined according to the state update validation rules as described in Section \ref{subsubsec:rules}.


\subsubsection{Functionality}
\label{subsubsec:basefunc}

\begin{itemize}
    
\item \textbf{Funding Phase}

\begin{figure}
    \centering
    \includegraphics[width=\linewidth]{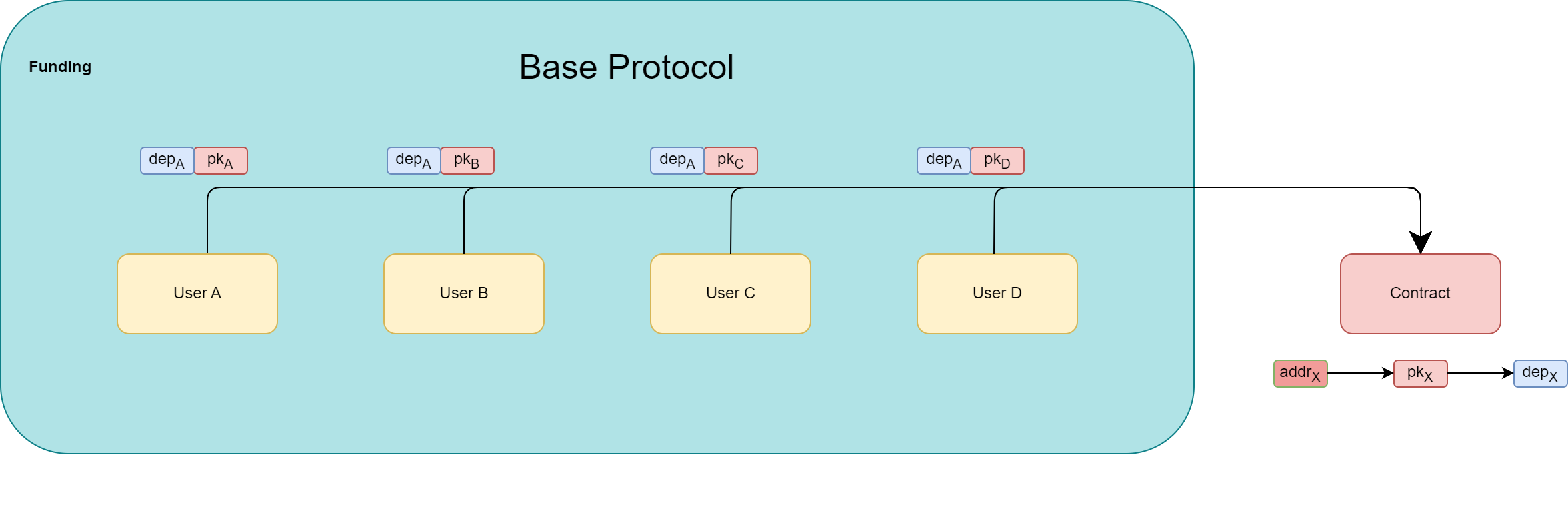}
    \caption{Depiction of the base channel funding process.}
    \label{fig:funding}
\end{figure}

Before the channel can begin its operation, every participant must provide their blockchain address and deposit funds that will be enough to function both as a stake and as funds to use in the process of running applications. Those values are stored in the Base Channel contract, mapped to each participant's address. The process is graphically represented in Figure \ref{fig:funding}. The record of a participant's funds deposit is a prerequisite for the corresponding state update that adds the user to the base channel balance sheet.

\item \textbf{Opening a new protocol}


A core feature of the Origami design is the recursive building of protocols on top of each other. The process of opening a new protocol on top of the current one is broken down in the following steps:

\textbf{Step 1:} Any channel participant that wants to create a new protocol can initiate this process. User A for this example, creates a new state of the Update: Open variety. To assemble this state, User A must fill in all the corresponding fields as those are defined in \ref{subssubsec:bcstate}. That includes:

\begin{enumerate}
    \item The creation of the new protocol's UCS through the following process:
    \begin{itemize}
        \item Setting the initial nonce to zero.
        \item Composing a Balance sheet that reflects the asset/fund commitments made by the members to the new protocol.
        \item Creating the new protocol's Channel ID by hashing the id of the base channel concatenated with the current state nonce for the base channel.
        \item Filling in the contract address that corresponds to the smart contract the new protocol is based on.
    \end{itemize}

\item User A must include in the state proposal the UCS Hash, and also add this hash to the Openings Accumulator of the base protocol.

\item The balance sheet of the current protocol must be updated to reflect that the funds committed to the new protocol are no longer spendable in this one.

\end{enumerate}
\textbf{Step 2:} User A, having created the proposal, must get it signed by the potential protocol members first. That means that the users who are included in the Balance Sheet for the new protocol must sign the state before it reaches the Header. At this point, the same users must also separately sign the new channel's UCS.

\textbf{Step 3:} The proposed state update can now be sent to the round's Header, who, along with the usual checks as those are described in Section \ref{subsubsec:rules}, must ensure the state has already been signed by all the potential members of the new protocol.

\textbf{Step 4:} The state is then propagated as described in Section \ref{subsubsec:comms}.

The process of opening a protocol is identical whether the protocol to be opened is a group or app protocol.

As the Opening of a group protocol does not happen on chain, it does not involve funding in the traditional sense. 

Members that wish to establish a group amongst themselves on top of the base protocol, propose a state update as it is defined in Section \ref{subsubsec:gopen}.




\item \textbf{Closing a Protocol}

App protocols have a finite purpose that once served calls for them to be closed, and at the same time they are the only protocols that closing applies to. The following steps describe the closing process that happens through a state update of the Update:Close variety in the base protocol:

\textbf{Step 1:} Any member of the protocol to be closed can commence this process. User A for this example is the one to create the closing state update which is comprised of:

\begin{enumerate}
    \item The Last Committed UCS that is currently representing the protocol in the Accumulator of Openings, whether that is UCS-0 or a more recent one through an Edit. 
    \item The hash of the aforementioned Last Committed UCS must also be included in the state.
    \item The most Recent UCS of the protocol, which is the one it is going to close on.
    \item The updated Accumulator of Openings from which the Last Committed UCS Hash representing the closing protocol has been removed.
    \item The balance sheet of the supporting protocol's UCS must be updated to reflect the redistribution of funds as they result from the most recent UCS of the closing protocol.
    \item The finalisability set that must be separately signed before it is appended to the state.

\end{enumerate}

\textbf{Step 2:} The update is sent to all members of the closing channel to be signed. The signed update is sent to the supporting channel's Header and the communication process proceeds as described in Section \ref{subsubsec:comms}.

\item \textbf{Editing a Protocol}

If Protocol B has been opened on top of the base protocol, and Protocol B's participant set changes, then the underlying protocol must be informed. This happens through a state update of the Update:Edit variety.

\textbf{Step 1:} Any member of the edited protocol can start off this process by creating the Edit state update in the supporting layer, in this case, the base protocol. This state update includes:

\begin{enumerate}
    \item The Last Committed UCS: UCS currently representing the protocol being edited in the accumulator of openings. If this is the first time the protocol is edited, then this is UCS0, otherwise it is the UCS that resulted from the previous edit.
    \item The hash of that Last Committed UCS
    \item The new UCS of the edited group that has the updated balance sheet and other protocol data, as well as its hash.
    \item The hash of the new UCS that will replace the Last Committed UCS Hash in the Accumulator.
    \item The Accumulator of Openings that has been updated by removing the Last Committed UCS and adding the New UCS in its place
\end{enumerate}

\textbf{Step 2:} The state update proposal is sent to the round's Header and the communication process resumes normally as described in \ref{subsubsec:comms}.

\item \textbf{Addition of participants}

\begin{figure}
    \centering
    \includegraphics[width=\linewidth]{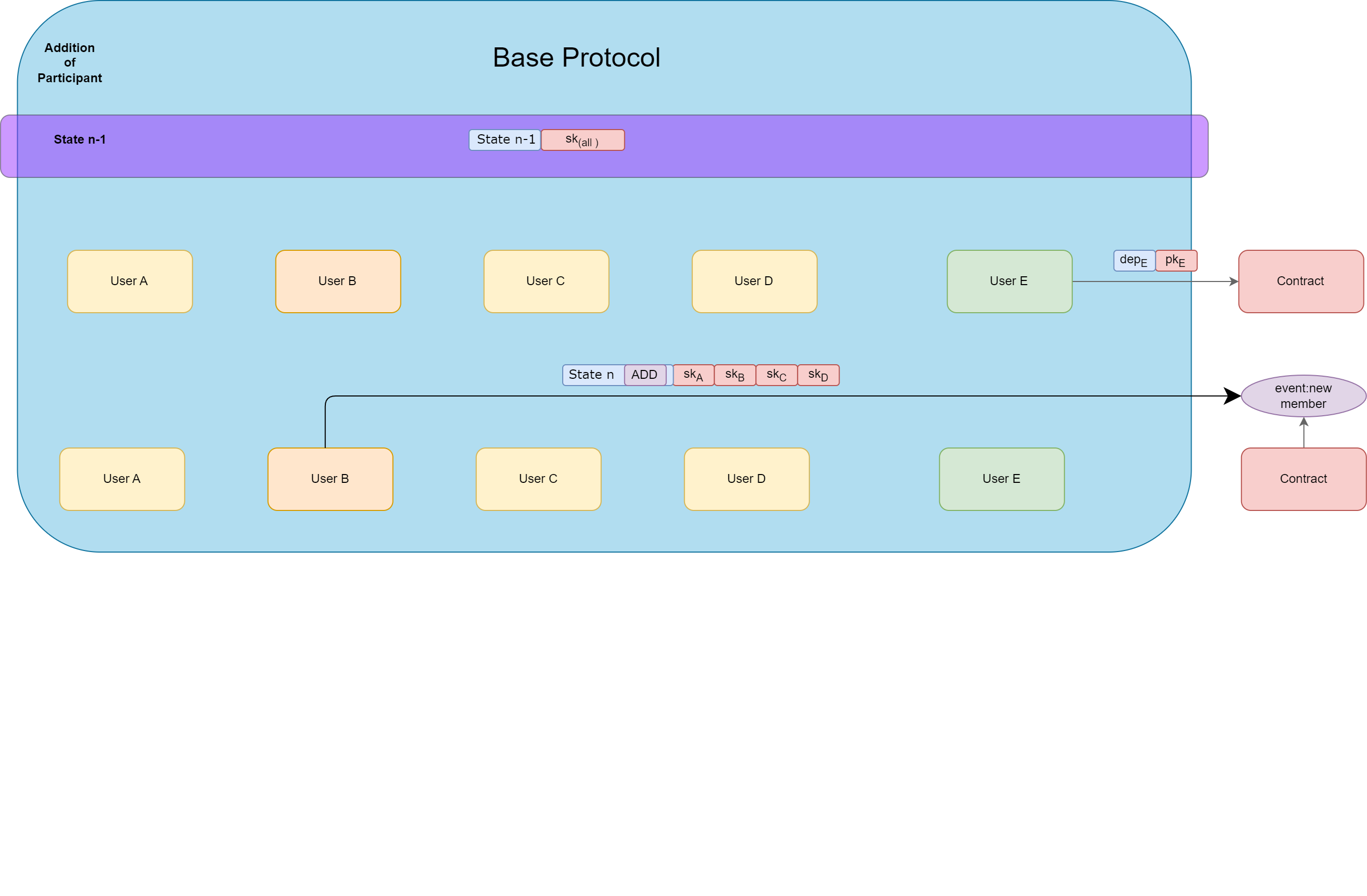}
    \caption{Base channel participant set expansion process.}
    \label{fig:bcexpansion}
\end{figure}

Origami allows the modification of the participant set of a channel after its creation. If a participant wants to join the base channel after it has been initially formed, then the following procedure, depicted in Figure \ref{fig:bcexpansion} is triggered. 

It is assumed the channel currently has 4 participants, users A through D, and user E wants to join.

\textbf{Step 1:} User E interacts with the channel's on-chain contract to deposit an amount of funds and this is recorded on the contract's storage. In this deposit, a small amount is included, dubbed as an "entrance fee", and functions as an incentive for the channels' participants to move along the process of including user E.

\textbf{Step 2:} The base channel contract emits an event to inform the participants that a joining request has been made by user E. 

\textbf{Step 3:} For user E to become a member, the cooperation of those already in the channel is required. Any user A through D can initiate the process of proposing a state update of type "Update:Add user", which includes the address of User E and defines the funds they committed to the contract and will be bringing into the channel.

\textbf{Step 4:} Given the fact that this state update is valid, it is signed by all channel members (including User E) and sets the new base channel state that includes E as a member. The user that initiated the update brings the new signed state update to the contract and the contract transfers to their account the entrance fee. The contract now has an updated member list and knows to take into account the signature of user E when judging whether a state is valid or not.

\item \textbf{Removal of participants}
\label{subsubsec:bgremove}

\begin{itemize}
    
\item \textbf{Upon Request:}
Any participant may decide they want to be removed from the base channel and retrieve their funds. 

\textbf{Step 1:} Any user has to prepare and propose an \textbf{Update:Remove User}, which removes themself or another user from the balance sheet.

\textbf{Step 2:} If the user proposing the update is not the user being removed from the channel, then the signature of the removed user must be obtained before the proposal is sent to the header.The user being removed from the balance sheet cannot be disputed against for inactivity, therefore no user can remove another from the balance sheet through this process without consent.

\textbf{Step 3:} When that state becomes valid, they can bring it to the contract to retrieve their funds. Through this process the contract is also made aware that the particular participant's signature is no longer needed.


\item \textbf{Unresolved Dispute:} It is common practice in state channel designs after an unresolved dispute to be forced to terminate the channel since a valid state can no longer be procured (not all members are present to sign). Wanting to eliminate this scenario, in Origami, the necessary on-chain interaction is taken advantage of in more than one way when an inactivity dispute remains unresolved:
The contract, after expiration of the time out, applies the inactivity penalty to the unresponsive party.Additionally, the contract will now stop requiring the inactive member's signature for a valid update. Therefore the channel can progress. 
\end{itemize}

\subsubsection{State update validation rules}
\label{subsubsec:rules}

The rules that define whether a state update proposal is valid are explicitly set and used by the members of the channel, as well as the state channel smart contract in case of a dispute. The process consists of two main parts:

\begin{enumerate}
    \item Ensuring the values of all fields are valid and correctly relate to the values in the previous state.
    \item Ensuring the state proposal always has the necessary signatures to move to the next step of the update process.
\end{enumerate}

\textbf{Open state update validation}

Given a \textbf{Open} state update is proposed by a channel member, the rest of the members shall validate the update. During this update a sub-set of the members of the channels are going to commit part of their funds to a new channel (group/app) based upon the base channel. The new state shall be validated according to the following rules:

\begin{itemize}
    \item \textbf{UCS:} Regarding the fields included in the UCS the necessary conditions are:
    \begin{itemize}
    \item \textbf{Nonce} value has to be increased by 1 with respect to the nonce of the previous state.
    \item \textbf{Balance sheet} shall be identical to the balance sheet of the previous state with the required modifications (balances decrease) according to the \textbf{Balance Sheet} in the \textbf{new UCS}.
    \item \textbf{Channel ID} must remain unchanged and correspond to the channel the update is taking place in.
    \item \textbf{Contract Address} has to correspond to the management contract of the protocol.
    \end{itemize}
    
    \item \textbf{New UCS:} Regarding the fields included in the new protocol's UCS the necessary conditions are:
    \begin{itemize}
    \item \textbf{Nonce} value has to be set to 0 with respect to the nonce of the previous state.
    \item \textbf{Balance sheet} shall include only the new channel's members and the funds they commit to the new protocol. Those funds must be equal to or less than their available funds in the base protocol.
    \item \textbf{Channel ID} is a new value produced by the hash value of the concatenation of the base channel id and the nonce value.
    \item \textbf{Contract Address} has to correspond to the management contract of the to-be formed protocol.
    \end{itemize}
    \item \textbf{New UCS Hash} is the hash of all the contents of the new protocol's UCS.
    
    \item \textbf{Accumulator of Openings} shall be the product of adding the new UCS Hash to the accumulator of the previous state.
    \item The \textbf{Header} value shall be calculated according to the mechanism defined in Subsection \ref{subsubsec:comms}.
\end{itemize}

In terms of necessary signatures, the conditions are the following:

\begin{itemize}
    \item \textbf{Condition 1} Before being sent to the header, the open state update proposal must have already been signed by all the members listed as participants for the new protocol that is to be opened.
    \item \textbf{Condition 2} Any member that is not to participate in the new channel must ensure that when they receive the update proposal, \textbf{Condition 1} is met, and additionally the header has also signed the proposal.
    \item \textbf{Condition 3} A valid state must carry a signature from every single member of the protocol.
    \item \textbf{Condition 4} The UCS must have been signed by every single member of the protocol it concerns before being appended to the state.
\end{itemize}

\textbf{Close state update validation}

A \textbf{Close} state update can be proposed by any member of an app protocol that wants to terminate, but must be validated by all members of the supporting channel before the protocol is permitted to close. Upon closing of the channel, the assets committed to it are returned to the supporting protocol and distributed according to the outcome as expressed by the closing state. The validity of the closing state is determined according to the following rules:

    \item \textbf{UCS:} Regarding the fields included in the UCS the necessary conditions are:
    \begin{itemize}
    \item \textbf{Nonce} value has to be increased by 1 with respect to the nonce of the previous state.
    \item \textbf{Balance sheet} shall be identical to the balance sheet of the previous state with the required modifications (balances increase) according to the \textbf{Balance Sheet} in the \textbf{Recent UCS}.
    \item \textbf{Channel ID} must remain unchanged and correspond to the channel the update is taking place in.
    \item \textbf{Contract Address} has to be a valid contract address that corresponds to the management contract of the protocol.
    \end{itemize}

    \item \textbf{Finalisability:} Regarding the fields included in the finalisability set, the necessary conditions are:
    \begin{itemize}
    \item \textbf{Finalisable Flag} must be set to true.
    \item \textbf{Nonce} value has to be the same as the nonce value included in the Recent UCS.
    \item \textbf{Channel ID} must remain unchanged and correspond to the channel that is to be closed.
    \end{itemize}
    
    \item \textbf{Last Committed UCS:} must include the data that produces the hash that currently represents the protocol to be closed in the Accumulator of Openings.
    \item \textbf{Last Committed UCS Hash:} must be identical to the element to be removed from the Accumulator of Openings, and the product of hashing the data contained in the Last Committed UCS.
    \item \textbf{Recent UCS} is the UCS identical to the one included in the most recent state update of the protocol to be closed that expresses the outcome of said protocol.
    \item \textbf{Accumulator of Openings} shall be the product of removing the new Last Committed UCS Hash from the accumulator of the previous state.
    \item The \textbf{Header} value shall be calculated according to the mechanism defined in Subsection \ref{subsubsec:comms}.
\end{itemize}

In terms of necessary signatures, the conditions are the following:

\begin{itemize}
    \item \textbf{Condition 1} Before being sent to the header, the close state update proposal must have already been signed by all the members listed as participants for the  protocol that is to be closed.
    \item \textbf{Condition 2} Any member that is not a participant of the closing channel must ensure that when they receive the update proposal, \textbf{Condition 1} is met, and additionally the header has also signed the proposal.
    \item \textbf{Condition 3} A valid state must carry a signature from every single member of the protocol.
    \item \textbf{Condition 4} The UCS must have been signed by every single member of the protocol it concerns before being appended to the state.
    \item \textbf{Condition 5} The Finalisability data set must have been signed by every single member of the protocol it concerns before being appended to the state.
\end{itemize}

\textbf{Edit state update validation}

\begin{itemize}
\item \textbf{UCS:} Regarding the fields included in the UCS the necessary conditions are:
    \begin{itemize}
    \item \textbf{Nonce} value has to be increased by 1 with respect to the nonce of the previous state.
    \item \textbf{Balance sheet} shall be identical to the balance sheet of the previous state with the required modifications (balances increase) according to the \textbf{Balance Sheet} in the \textbf{Recent UCS}.
    \item \textbf{Channel ID} must remain unchanged and correspond to the channel the update is taking place in.
    \item \textbf{Contract Address} has to be a valid contract address that corresponds to the management contract of the protocol.
    \end{itemize}
    
    \item \textbf{Last Committed UCS:} must include the data that produces the hash that currently represents the protocol to be closed in the Accumulator of Openings.
    \item \textbf{Last Committed UCS Hash:} must be identical to the element to be removed from the Accumulator of Openings, and the product of hashing the data contained in the Last Committed UCS.
    \item \textbf{New UCS} is the UCS that includes the changes in the member list and fund distribution of the edited channel and will replace the current UCS representing it in the Accumulator of Openings.
     \item \textbf{New UCS Hash} is the hash of the data included in the new UCS and the element to be added in the Accumulator of Openings.
    \item \textbf{Accumulator of Openings} shall be the product of removing the new Last Committed UCS Hash from the accumulator of the previous state and replacing it with the New UCS Hash.
    \item The \textbf{Header} value shall be calculated according to the mechanism defined in Subsection \ref{subsubsec:comms}.
\end{itemize}

\begin{itemize}
    \item \textbf{Condition 1} Every member must ensure that when they receive the edit update proposal, it has already been signed by the header.
    \item \textbf{Condition 2} A valid state must carry a signature from every single member of the protocol.
    \item \textbf{Condition 3} The UCS must have been signed by every single member of the protocol it concerns before being appended to the state.
\end{itemize}

\textbf{Add user state update validation}

Given an Add user state update is proposed by a channel member, the rest of the members shall validate the update. The new state shall be validated according to the following rules:

\begin{itemize}
    \item \textbf{UCS:} Regarding the fields included in the UCS the necessary conditions are:
    \begin{itemize}
    \item \textbf{Nonce} value has to be increased by 1 with respect to the nonce of the previous state.
    \item \textbf{Balance sheet} shall be identical to the balance sheet of the previous state with two exceptions. Firstly, the addition of an entry with the new user and the corresponding balance, according to the deposit in the base channel contract. Additionally, the payment of the entrance fee from the new member to the proposer of the state must be reflected in the balance sheet.
    \item \textbf{Channel ID} must remain unchanged and correspond to the channel the update is taking place in.
    \item \textbf{Contract Address} has to correspond to the management contract of the protocol. (Origami contract for base and group protocols, respective application contract for app protocols) 
    \end{itemize}
    \item The  values in the sub-fields \textbf{User address} and \textbf{Amount} of the \textbf{Update} field must correspond to the new entry and fund addition reflected in the updated balance sheet.
    \item \textbf{Accumulator of Openings} shall be identical to the one of the previous state.
    \item \textbf{Header} value shall be calculated according to the mechanism defined in Subsection \ref{subsubsec:comms}.
\end{itemize}

\begin{itemize}
    \item \textbf{Condition 1} Every member must ensure that when they receive the edit update proposal, it has already been signed by the header.
    \item \textbf{Condition 2} A valid state must carry a signature from every single member of the protocol, including the new user being added.
    \item \textbf{Condition 3} The UCS must have been signed by every single member of the protocol it concerns before being appended to the state.
\end{itemize}

\textbf{Remove user state update validation}

Given a \textbf{Remove user} state update is proposed by a channel member, the rest of the members shall validate the update. The new state shall be validated according to the following rules:

\begin{itemize}
    \item \textbf{UCS:} Regarding the fields included in the UCS the necessary conditions are:
    \begin{itemize}
    \item \textbf{Nonce} value has to be increased by 1 with respect to the nonce of the previous state
    \item \textbf{Balance sheet} shall be identical to the balance sheet of the previous state with the exception of the removal of the entry of the user to be removed. 
    \item \textbf{Channel ID} must remain unchanged and correspond to the channel the update is taking place in.
    \item \textbf{Contract Address} has to correspond to the management contract of the protocol. (Origami contract for base and group protocols, respective application contract for app protocols)
    \end{itemize}
    \item The values in the sub-fields \textbf{User address} and \textbf{Amount} of the \textbf{Update} field shall be set to the address of the user to be removed and the balance they had in the previous state.
    \item \textbf{Accumulator of Openings} shall be identical to the one of the previous state.
    \item \textbf{Header} value shall be calculated according to the mechanism defined in Subsection \ref{subsubsec:comms}.
\end{itemize}

\begin{itemize}
    \item \textbf{Condition 1} Every member must ensure that when they receive the edit update proposal, it has already been signed by the header.
    \item \textbf{Condition 2} A valid state must carry a signature from every single member of the protocol, including the user that is being removed.
    \item \textbf{Condition 3} The UCS must have been signed by every single member of the protocol it concerns before being appended to the state.
\end{itemize}

\subsection{Group Protocol}
\label{ssubsec:basegroup}
Origami is meant to remain active indefinitely. However, the addition of participants to the base protocol in a boundless manner would undoubtedly result to high latency, eventually rendering the channel unusable.

The group protocol has been designed as a scaling mechanism to prevent this scenario. Subsets of base protocol members create these instances that can stack infinitely on top of each-other, limiting the number of signatures required for a state update. The goal is that as the number of Origami participants escalates, the system scales upwards and the lower, densely populated protocols are rarely active.

The group protocol shares most of its functionality with the base protocol, since it is based on the same smart contract and serves the same purpose. group protocols can support group or app protocols opening on top of them. 

Specifically, the following processes are identical between the two protocols:

\begin{itemize}
    \item \textit{Representation of the State:} The varieties and contents of the channel states are shared between the base and group protocols.
    \item \textit{Communication model:}Since group protocols are governed by the same smart contract as the base protocol, they also follow the same, non-turn based communication scheme that is described in Section \ref{subsubsec:comms}.
    \item \textit{Disputes:} Disputes are also handled by the same contract and therefore in the same manner.
    \item \textit{Opening, Closing and Editing another Protocol functionalities: The manner in which a channel is opened, closed or edited does not differ between it happening on top of a base protocol or a group protocol.}
\end{itemize}

The present section will be elaborating on the functionalities that differ between the base and group protocols.

\subsubsection{Group Protocol Opening} 
\label{subsubsec:gopen}

For the opening of a group protocol, actions in two layers are necessary. Part of the process happens in the supporting channel (the base or group protocol underneath) and part of it in the new channel that is being opened. Actions in the supporting channel are those described in Section \ref{ssubsec:basegroup}: \textbf{Opening a new protocol}. In this section the actions taken in the new channel during its creation are analysed.

After establishing the new group protocol in the supporting channel, the members must adhere to the following process:

\textbf{Step 1:} A starting state must be propagated. This state includes:
\begin{itemize}
    \item The new channel's UCS that was signed by the members during the opening of the group as described in Section \ref{subsubsec:basefunc}.
    \item An empty Accumulator of Openings.
    \item The header for the next round defined as the first entry on the balance sheet.
\end{itemize}

The creation of the starting state is done by the protocol member that appears first on the balance sheet, for coordination reasons.

\subsubsection{Addition of Participant}
\label{gpadd}

Group protocols also allow the modification of their participant set. However, the process differs fundamentally from the base protocol one, since no on-chain action takes place to alert participants of the to-be member's request.

The addition of a participant to an existing group protocol is modeled as such:

\textbf{Step 1:} It is assumed that the group protocol has 4 participants, Users A through D, and User E aims to join. User E communicates their entry request to any member of the group protocol.

\textbf{Step 2:} The group protocol member that adds User E, lets say User A, must create two state updates:
    \begin{itemize}
        \item \textit{Group Protocol State Update:} This update is propagated inside the group protocol. It adds User E and their funds to the protocol's balance sheet, and also reflects the payment of the entrance fee from User E to User A. This state update is what is defined as an \textbf{Add User} type of update in Section \ref{subssubsec:bcstate}.
        \item \textit{Supporting Layer State Update:} Any group protocol has an underlying layer that must also be updated after any change to the participant set. The new state is meant to replace the original opening state of the group protocol, showing the new participant set and the new balance sheet. This is necessary to keep all participants as well as the smart contract aware of the alterations. This state update is what is defined as an \textbf{Edit} type of update in Section \ref{subssubsec:bcstate}.
    \end{itemize}

\subsubsection{Removal of Participant}
\label{gpremove}
A participant's removal from a group protocol is called upon in the same cases as the base protocol, but are handled differently because the do not equate removal from the ecosystem.

\textbf{Upon Request:} A participant that wants to voluntarily leave a group protocol they are in has to create two new states, a \textit{Group Protocol State Update} of the \textbf{Remove} variety, and an {Underlying Layer state update} of the \textbf{Edit} variety. Those serve the same purposes and have the same contents as described in Section \ref{gpadd} but now indicate a participant's removal and the resulting transferring of funds from the group protocol to its underlying layer.
 
\textbf{Unresolved Inactivity Dispute:} To prevent the stagnation of the protocol processes in the event of an unresolved Inactivity Dispute, the party at fault is removed. The necessary state updates in both the group protocol and underlying layer will lack the party at fault's signature, and therefore the unresolved state of the dispute must be visible in the contract to make those updates valid.


\subsection{App Protocol}

The app protocol maintains the purpose of a traditional state channel, which is to run a blockchain application off-chain while only involving directly interested parties. App protocols are opened on top of a group or base protocol any time a subset of participants wish to run a state channel application amongst themselves.

The ability to run any application amongst any combination of origami participants can eliminate the need to ever form another state channel, therefore making this scheme very convenient and also reducing channel creation costs and transaction load on the blockchain.


Due to being higher layers in the origami protocol, app protocols share a number of functionalities with group protocols. Those are:

\textbf{Opening:Opening an app protocol happens through the same process as a group protocol, by requiring a state update of the \textbf{Open} variety to take place in the supporting protocol. The functionality within the new channel is defined by the smart contract of the application.}
\textbf{Participant set Modification:}The option to modify the participant set is left upon the designers of the application smart contract an app protocol is running, because it may or may not fit into the nature of the application. If desired, however, the application smart contract can incorporate the modification of its participant set as described in Sections \ref{gpadd} and \ref{gpremove}.

The rest of this section will elaborate on the functionalities unique to app protocols.





\subsubsection{App Protocol Communication}

In-group communication follows whatever pattern the application developers have designed. The non-turn based communication protocol described in Section \ref{subsubsec:comms} can be incorporated into an application, but it is not necessary in case turn-based communication is preferred.

\subsubsection{App Protocol Inactivity Dispute}
\label{subsec:appdispute}

An inactivity dispute for an app protocol is triggered through the application's contract. Because the contract is not contacted unless there is misbehaviour, it is not aware of the members of the group or any other specific information. Therefore, the data included in the app protocol's opening state is provided to the application contract when a dispute is initiated. How the dispute is resolved also depends on the smart contract specific to the application. 


\subsubsection{App Protocol Closing}
\label{subsubsec:appclose}

\begin{figure}
    \centering
    \includegraphics[width=\linewidth]{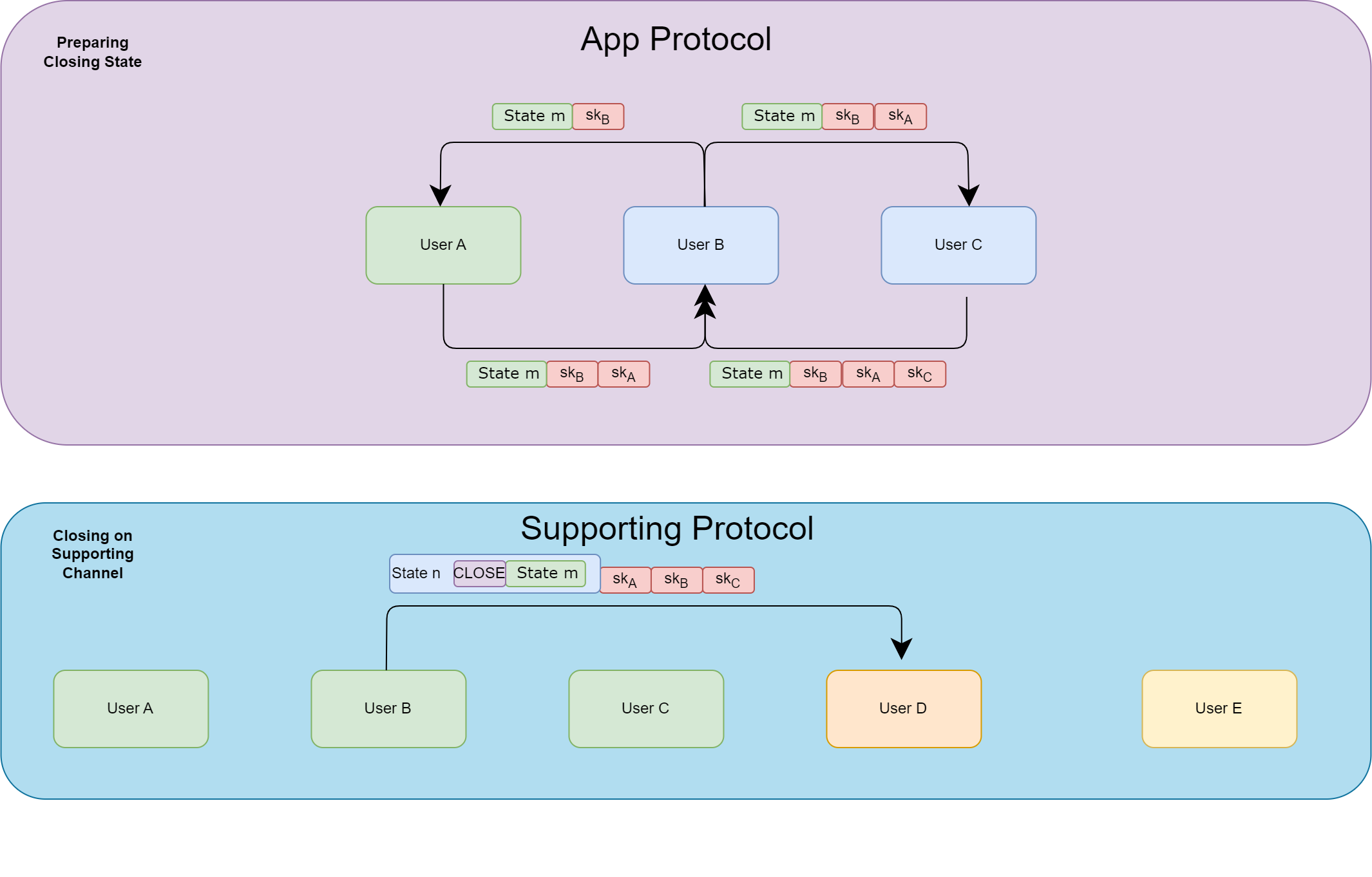}
    \caption{App Protocol closing process.}
    \label{fig:subgroupclose}
\end{figure}

When the application run in the app protocol has fulfilled its purpose, the participants can close the channel and release the funds locked in it back in the underlying channel. The process to close an app protocol is described by Figure \ref{fig:subgroupclose} and detailed in the following steps:

\begin{itemize}
    \item \textbf{Step 1} Any member of the app protocol can initiate the process of closing it as long as the most recent state of the protocol is a finalisable state, which is a state appropriate to close upon. Finalisability is determined by the individual app contract and signaled through the independently signed Finalisable set as this is defined in Section \ref{subssubsec:bcstate}. 
    \item \textbf{Step 2} Closing an app protocol is synonymous with informing the channel it was opened on of its dissolution. The channel members must prepare and sign a closing state for the supporting base or group channel that is of the \textbf{Close} variety that will, among other values, include the UCS of the state to close upon and the Finalisable set in the Update field that will allow the Origami management contract to determine the validity of the state update proposal. 
\end{itemize}


\section{Protocol Relationships}

\begin{figure}
    \centering
    \includegraphics[width=\linewidth]{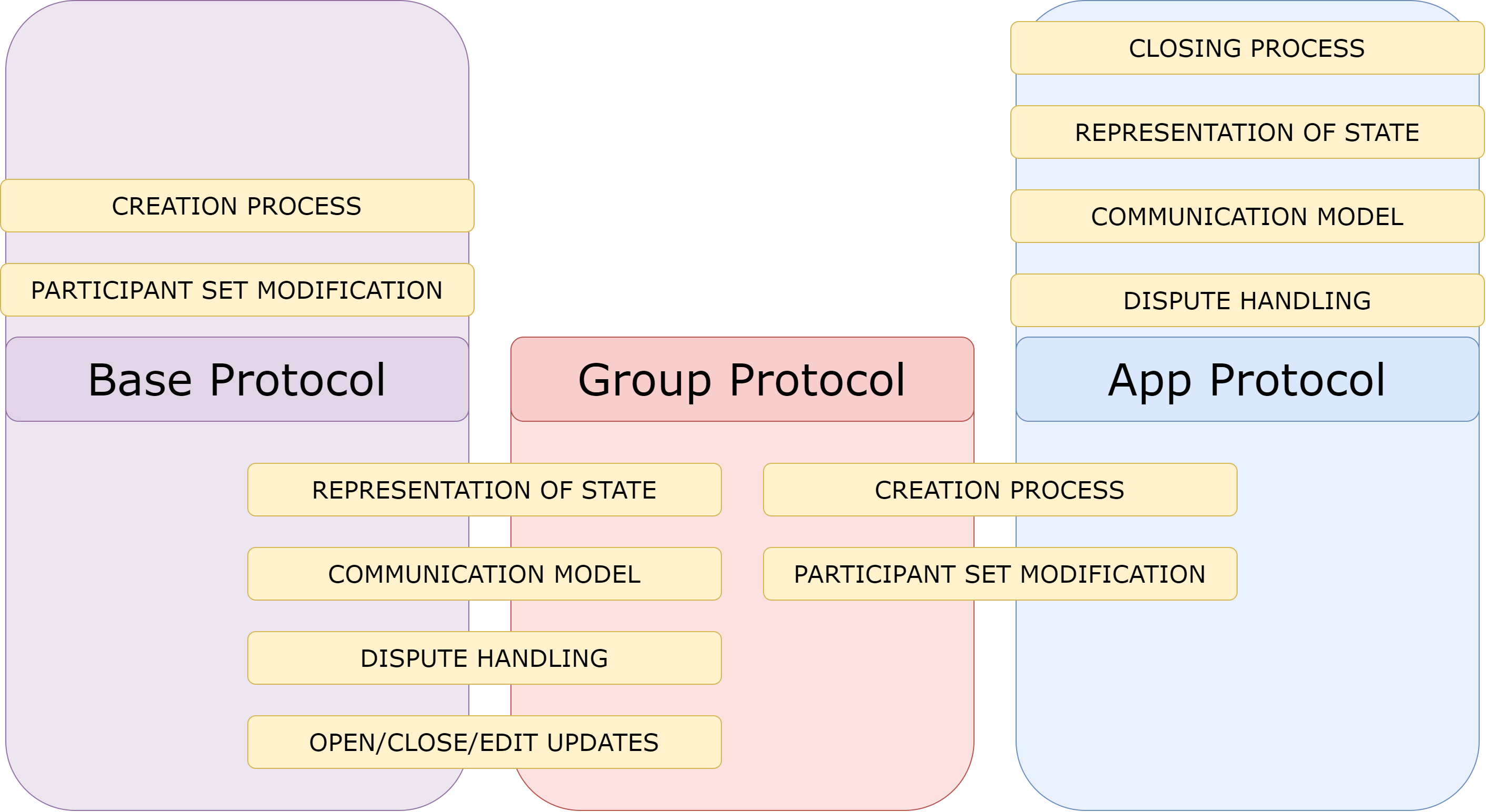}
    \caption{Functionalities per protocol.}
    \label{fig:sharedfunc}
\end{figure}

In Figure \ref{fig:sharedfunc} an overview of all Origami functionalities and how those are divided per protocol can be seen.

The base protocol has a unique creation process and way of handling modifications in the participant set. This stems from the fact that, as the entryway to the Origami ecosystem, interacting with the base channel sometimes requires interacting with the blockchain network.

App protocols are the only protocols where closing is applicable as app channels most often have a finite purpose. The communication model an app protocol uses can either have a more traditional turn-based approach or employ the communication scheme introduced by Origami in Section \ref{subsubsec:comms}, depending on the needs of the particular application. Dispute handling is also case-specific and determined by the application smart contract. The same stands for representation of the state with the exception of being required to include the Finalisability set as that is described in Section \ref{subssubsec:bcstate}.

Group protocols are distinct in the sense that they do not have any unique functionalities. The communication model employed is the one analysed in Section \ref{subsubsec:comms}, which also holds true for the base protocol. Dispute handling as well as the representation of the state are also shared between base and group protocols.
Since both group and app protocols are higher layers that only require on-chain interaction for disputes, they have the same creation process and implement participant set modification in identical ways.

\subsection{Origami Smart Contracts}
\label{subsec:contracts}

Every state channel is managed by a smart contract that governs its funds and settles its disputes. The base and group protocols share a single management contract, called the Origami Management Contract. Each app protocol though corresponds to a state channel application contract that governs the specific application which are dubbed Origami app contracts.

Apps that aim to be compatible with the Origami design must make minimal modifications to their smart contract in order to be able to take advantage of its functionalities. Origami app contracts must therefore meet the following conditions:

\begin{itemize}
    \item Conventional state channel contracts are not usually usable until the funding phase has been complete. In contrast, Origami app contracts are not meant to be interacted with unless a dispute occurs, and even in that case, they do not accept disputes only by a single channel, but by any app protocol that is running the particular application. Therefore, instead of a funding phase that sets the list of participants, Origami apps must accept the opening state information that is unique for each app protocol.
    \item In order to enable the handling of disputes the lower levels, the Universal Channel State has been defined as an independently signed data set that enables the flow of channel data between app, group and base protocols. Origami app contracts must take care to include this separate set along with their state updates.
    \item Finally, every Origami App state must include the independently signed flag set that includes the flag, the channel ID and the channel nonce and indicates whether a state is finalisable or not, as described in Section \ref{subssubsec:bcstate}. The app smart contract must also check whether the value of the flag is appropriate for the specific state (eg. if the state whose finalisable flag value is true is indeed finalisable).
\end{itemize}

\section{Comparative Evaluation}
\label{sec:evaluation}

In a previous work \cite{negka2021blockchain}, a comparative evaluation of the majority of state channel designs was performed. This effort helped identify the common shortcomings of works in the state channel research field, as well as the more desirable characteristics. Table \ref{table:comparison} is the graphical representation of the results of this analysis. Table \ref{table:load} is an effort to measure the transaction load each state channel design brings to the blockchain and as a result, evaluate how well the design has served its purpose of decreasing the number of on-chain transactions.

The main motivation behind Origami was creating a design with minimum negative characteristics and maximum beneficial features, while also maintaining a low number of necessary on-chain transactions. In this section, Origami is evaluated alongside other state channel designs across the same measures set in \cite{negka2021blockchain}.

\subsection{Requirements}

\begin{figure}
    \centering
    \includegraphics[width=\linewidth]{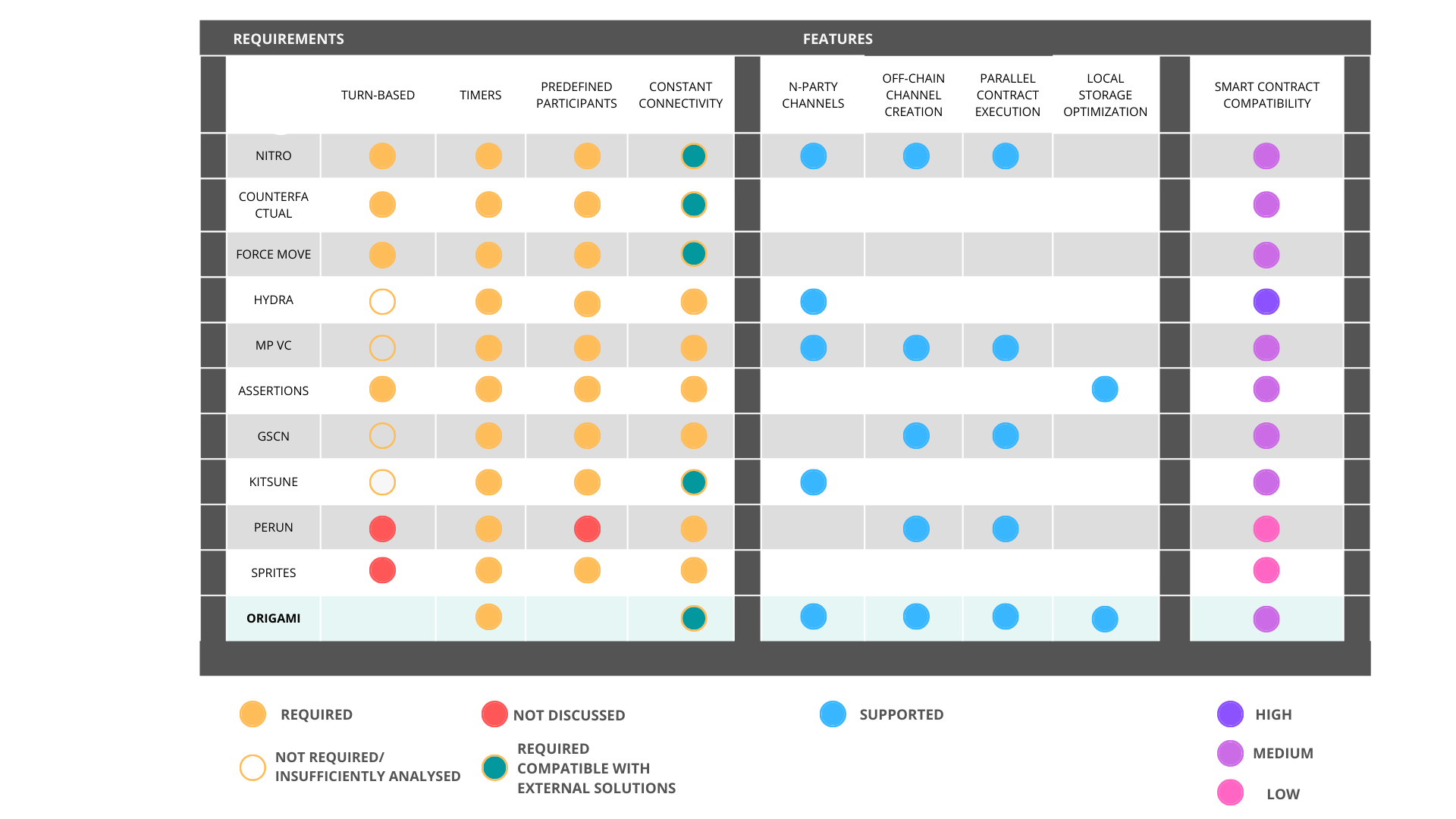}
    \caption{Comparative evaluation of Origami and other state channel designs.}
    \label{table:comparison}
\end{figure}

The first section of Table \ref{table:comparison} titled "Requirements" is focused on restrictions state channel designs place on applications in order to support them. A higher number of requirements means a more limited pool of applications is compatible with the design, while some of the demands decrease usability. Therefore, the fewer cells checked on the Requirement section, the broader the scope of applications a design can support.

\begin{itemize}

\item \textbf{Turn-based Applications:}
Limiting the nature of supported applications to ones that function with turn-based communication is a matter of accountability distribution. Stale update posters and inactive members are more easily identified if the order of updating is predefined. As a result, it is common for designs to enforce this restriction on the apps they support.
While a few of the analysed schemes do not explicitly place the turn-based requirement, Origami is the only one that extensively elaborates on how non-turn based communication is realised and how concurrent update proposals would be handled, as analysed in Section \ref{subsubsec:comms}.

\item \textbf{Timers:}
Disputes in state channels are structured as a challenge-response mechanism. A challenge is issued and there must be a response before the time limit goes by, both for potential contentions and for a successful resolution. Because there is no provable way to both securely and accurately measure time on the blockchain, this is not a preferred practice. However, as no workaround to this method has been developed, this is a requirement shared by all state channel designs.

\item \textbf{Predefined Participant Sets:}
To avoid complicating the funding process, and also limit interactions with the smart contract, state channel designs enforce static participant sets. This further enforces the need to create a separate state channel if there is desire for an addition to the initial set, and the need to close the channel when a participant wishes to leave.
In contrast, Origami allows total flexibility in the participants of the base channel or any of the higher levels. Mechanisms to join and leave on demand are outlined in Sections \ref{subsec:basechannel} and \ref{ssubsec:basegroup}. The only on-chain transaction in the whole process is funding the base channel, a one time requirement to then freely enter, form and leave any higher layer.
Unresolved disputes also result in member "expulsions" to avoid forcing the group or channel to close when singular members become unavailable or misbehave.

\item \textbf{Constant Connectivity:}
Inactivity disputes are integral to the state channel functionality. They apply penalties to any member that obstructs state progression with their absence. The repercussions are applied regardless of whether the absence is the result of malicious behaviour or inability to respond. Therefore, by extent, channel members cannot risk being unavailable under any circumstances, or they stand to suffer consequences or lose their chance to contest a dispute.
Designs like Pisa \cite{mccorry2019pisa} and Brick \cite{avarikioti2019brick} are solutions to the state channel design that allow some form of delegation of signatures from a member that intends to be absent to an external entity. However, these solutions either require additional smart contracts to be set up or compromise the privacy of the channel. Additionally, they are not compatible with all state channel designs.

It must be noted that because \cite{mccorry2019pisa} and \cite{avarikioti2019brick} are efforts of limited scope and aim to limit only this specific requirement, they cannot be fairly compared to other solutions and have been omitted from comparison in Table \ref{table:comparison}. 

\end{itemize}

\subsection{Features}

The second section of Table \ref{table:comparison} titled "Features" lists a number of desirable functionalities a state channel design can offer. Whether its an increase in efficiency, usability, or a reduction in cost, multiple checks in this section elevate the design that can boast them. 

\begin{itemize}

\item \textbf{Multi-party Channels:} The earlier state channel designs were built for two party interaction. There have been many successful attempts to overcome this limitation ever since with a number of designs offering the option to build channel between an unlimited number of users, as long as they belong in a predefined set. Origami also places no restriction on the number of members in a channel or group, and even allows the participant set composition to be altered after channel establishment.

\item \textbf{Off-chain Channel Creation:} In order to further limit the transaction load on the blockchain, attempts were made to enable the creation of state channels without any interaction with the chain. The result of these efforts were Virtual Channels, or otherwise called State Channel Networks (\cite{close2019nitro}, \cite{dziembowski2018general}, \cite{dziembowski2019multi}). The primary feature of such designs is that they allow parties to form state channels off-chain, as long as they have each already formed one with a common intermediary. This can also be extended, providing the ability to form channels off-chain as long as there is any path of preformed state channels that can connect participants.
The drawbacks of those designs stem from the necessity that an intermediary exists. Given Users A,B and C, and assuming B will be acting as intermediary between A and C, it is necessary that B has enough funds committed in their state channel with A to cover C's bond, and vise versa. Essentially, User B's funds are now locked and unusable. On top of that, the funds either A or C can use are limited to the amount B has in their respective channels. This limitation increases in severity if the path between A and C is even more extended.
Origami creates an ecosystem where participants can open and close groups that function as channels without on-chain interaction. However, since there is no requirement for an intermediary, Users are only limited by their own deposit, and nobody has to lock their funds to enable the communication of third parties.


\item \textbf{Smart Contract Compatibility:} This column refers to the level of modifications that need to be made in an application in order to make it compatible with a state channel design. Low compatibility requires for a smart contract to be redesigned from scratch in order to be able to run on a state channel. Medium compatibility refers to most cases, where a contract is channel-compatible after a few modifications (implementation of a specific function, use of a specific library). High compatibility is only achieved by \cite{chakravarty2020hydra} that allows for the use of a contract in a state channel with zero modifications, thanks to the use of the EUTXO model that is Bitcoin-specific. 

\item \textbf{Parallel Contract Processing:} A channel that can process more than a single contract at a time can also support more than a single application running at a time. For state channel networks this means more than a single virtual channel can be supported by any given channel at a time. In Origami, there is no limit on how many app protocols can be run at one time, since they do not require any kind of contract instantiation.


\item \textbf{Local Storage Optimisation:} By making use of cryptographic accumulators, Origami severely cuts down on the local storage demands its states impose on its users. Additionally, the design eliminates the need to be maintaining any more than the most recent state for each channel a user participates in and the opening state of that channel.

\end{itemize}

\subsection{Efficiency}

\begin{figure}
    \centering
    \includegraphics[width=\linewidth]{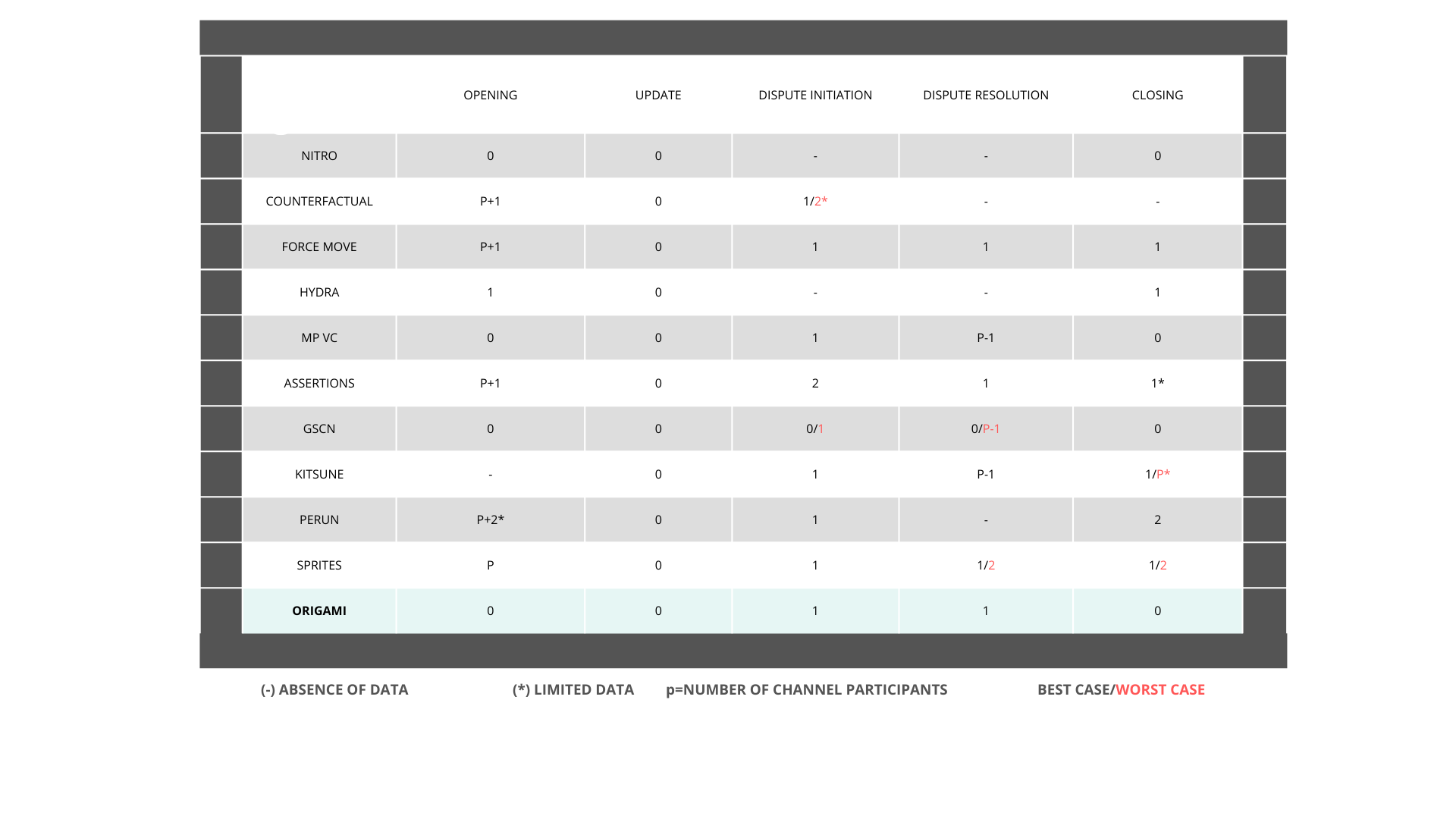}
    \caption{Comparison of transaction load per phase of channel.}
    \label{table:load}
\end{figure}

State channels were developed with the primary intent of decreasing the transaction load on the blockchain, and with it the fees associated with on-chain transactions. Measuring transaction fees with any accuracy is impossible since there is a great dependency on the nature and state of the application contract, the blockchain system and the greatly fluctuating cost of gas units.
Measuring the quantity of on-chain transactions per state channel phase is possible, and while not a perfectly comparison, the numbers provide a decent indication of a design's efficiency. Such an evaluation can be seen in Table \ref{table:load}, with the phases being defined as Opening, Update, Dispute Initiation, Dispute Resolution and Closing.

\textbf{Opening:} A typical state channel opening process includes the deployment of the channel contract and its funding. There is some differentiation in the funding process between designs.
Creating a virtual channel requires no on-chain interaction, but it does require that ledger channels already exist, at least as many as the virtual channel's participants. When marking the opening transactions required as 0, this refers to the creation of the virtual channel and does not take into account the transactions needed for the ledger channels that are preexisting.
By the same logic, Origami groups require 0 on-chain transactions to be formed once in the ecosystem. In stark contrast to virtual channels though, the single transaction required to join the ecosystem guarantees any subsequent groups will be formed off-chain, regardless of whether there is a direct connection with other participants or not.

\textbf{Update:} Since state channels exist to enable off-chain state progression, it is to be expected that in the optimistic case there will be no communication with the blockchain during the update phase.

\textbf{Dispute:} Even though this is the phase where designs diverge the most from one another, it is most often in the form of the dispute transaction submitted rather than the number of transactions necessary. To better analyse the Dispute phase we have split it into two sub-phases on Table \ref{table:load}, Dispute Initiation and Dispute Resolution.

\textbf{Closing:} Like the opening phase, closing also follows a standard pattern, that of redistributing assets to the channel's participants. Whether that distribution happens on-chain or off-chain is what sets designs apart in this phase. Conventional state channel designs perform the distribution on chain and usually demand as many interactions with the contract as there are participants. Virtual Channels and Origami allow this process to happen off-chain.

Because the amount of data available for the construction of Table \ref{table:load} was abundant in some cases while limited to non-existent in others, there has been an effort to mark the availability of information to draw from for each evaluated paper.

\section{ Universal Composability }
\label{sec:security}
In our model, following the work of \cite{dziembowski2018general}, which is inspired by the universal composability framework of Canetti \cite{canetti2005security} and uses a synchronous version of the global UC framework(GUC), we build a UC-style model that utilizes many features, assumptions, and set-ups from their work, in order to build a protocol that guarantees our security goals. In the following paragraphs \ref{sec:definitions} we introduce the definitions and syntax  used to describe the protocols that form the ideal functionality and the adversarial model. We assume that the parties that are connected by authentic communication channels execute a $n-party$ protocol $\Pi$ in the presence of an adversary $Adv$ and the $environment Z$ represents everything external to the protocol and has access to the ideal functionalities. Then the ideal functionality is presented \ref{subsec:idealfunctionality} followed by the protocol \ref{sec:protocol} that describes analytically the actions of every member, which are observed as outputs of the protocol by the $environment Z$.  In \ref{subsec:security} the main security goals of our model are presented with regard to the adversarial behaviour that is presented in the last section of the simulator $Sim$ \ref{subsec:simulator}

\subsection{ Definitions }
\label{sec:definitions}

In the following section, we describe the ideal functionality that defines how the origami ecosystem is built and maintained but firstly we present the basic definitions and assumptions of the proposed model. Origami is implemented using a smart contract upon which manifold channels can be established and operated. In \cite{dziembowski2018general} a smart contract is formally defined as a tuple \( C = (\Lambda, g_1,...,g_r, f_1,...,f_s)\) , where \(\Lambda\) are the admissible contract storages and \(g_1,...,g_r\) are functions called contract constructors and \(f_1,...,f_s\) are called contract functions. Each contract constructor \(g_i\) is a function that takes as input a tuple \(\ P ( P, \tau, z \), with \(\ P \in \mathit{P}\), \(\tau\in N\) and \(\ z \in {0,1}\) and produces as output an admissible contract storage \(\sigma\) or a special symbol \(\bot\) that denotes that the contract construction failed. In this definition, \(P \) is the party that called the function, \(\tau\) the current round and \(z\) is used to pass additional parameters to \(g_i\). The constructors are used to create a new instance of the contract. If the contract construction did not fail, then \( g_i( P, \tau, z )\) is the initial storage of a new contract instance.\cite{dziembowski2018general} 

The channel created within the origami ecosystem can be formally presented as a tuple: $b:= (b.nonce, b.balance[b.members], b.id, b.contract,b.accumulator, b.header, b.update, b.signature) $, where \(b.members \in \mathit{P}\), \(\forall \mathit{P_{i}} \in \mathit{P} \) there is an account on the ledger and the money associated with each account are handled via a global ideal functionality \(\mathit{L}\) , formally presented in the work of \cite{dziembowski2018general}. B.balance is a mapping function that associates every member to the amount they deposited in the channel and $b.nonce$, $b.id$, $b.contract$, $b.accumulator$,$b.header$ and $b.update$ correspond to the fields presented in \ref{subsec:basechannel}. Finally, yet importantly the attribute \(b.signature\) represents the field where the signatures of every member in \(b.members \in \mathit{P}\) are stored in every update of the state of the channel.

In our model, the contract uses \(D\) and \(R\) storages, \(D,R \in \Lambda \) where information about deposits and refunds are stored, and the function validate-state-transition \(f_v\) that takes us input \( (S_b, S_b') \), where \( S_b'\) the state of the channel with \(b'.nonce= b.nonce + 1\) and outputs whether the state transition is valid.

\subsection{ Adding a participant} 
In this section, in order to confine the length of the paper, we focus our security analysis on the base group model and especially the function of adding a new member to it. The ideal functionality describes the addition of a new member, followed by the presentation of the protocol where the actions of every involved party are thoroughly explained. 
Let \(\mathit{P_N}\in \mathit{P}\) the new member that wants to join the base channel \(b\) locking \(\mathit{c,f}\) coins into it, as described in \ref{subsec:basechannel} a predefined amount of coins \(f\)is used as an entrance fee that compensates the member \(\mathit{P_x}\) that includes the new participant \(\mathit{P_N}\) in a future state update \(b'\). Additionally, let \(\mathit{P_s}= P\{-\mathit{P_x}\}\) the rest of the users of the base channel whose consensus is necessary for the state update. 

\subsubsection{ Ideal Functionality $F_{bch}$ - Adding a participant} 
\label{subsec:idealfunctionality}
\begin{tcolorbox}[title=Ideal Functionality \(\mathit{F_{bch}}\),colback=white]

\begin{tcolorbox}
Add participant
\end{tcolorbox}
Upon \( (add, \ P , b ) \xleftarrow{\text{\(\tau_0 \)}}\) \ P :
\begin{enumerate}
\item \( (deposit, \ P , b , c , f ) \xleftarrow{\text{\(\tau_0 \)}}\) \ P
\item If after \(\Delta\) rounds \(\ P \notin b.members \)\\
Upon \( (refund, \ P , b , c , f ) \xleftarrow{\text{\(\tau_0 +\Delta  \)}}\) \ P\\
add \(c + f\) coins to \(\ P's\)  account on \( \Hat{L} \)\\
\end{enumerate}

\end{tcolorbox}

\subsubsection{ Protocol  $ \Pi $  }
\label{sec:protocol}

\begin{itemize}

\item  Party \(\mathit{P_N}\) upon \((add, P_N, b, c, f) \xleftarrow{\text{\(\tau_0 \)}} \mathit{Z}\)\\
 1. Send \( (deposit,P_N, b, c, f) \xrightarrow{\text{\(\tau_0 \)}} \mathit{F_{bch}}\)
\item Party \(\mathit{P_x}\) upon \((add, P_N, b, c, f) \xleftarrow{\text{\(\tau_0 \)}} \mathit{Z}\)\\
2. If \( (add\_request,P_N, b, c, f) \xleftarrow{\text{\(\tau_1 = \tau_0 + \mathit{\Delta}\)}} \mathit{F_{bch}}\) then \\
\( (add-user,P_N, b, c, f) \xrightarrow{\text{\(\tau_2\)}} \mathit{F_{bch}}\)
\item Back to Party \(\mathit{P_N}\)\\
3. If \( (user\_added,P_N, b, c, f, \mathit{P_x}) \xleftarrow{\text{\(\tau_2 \)}} \mathit{F_{bch}}\) and if $d.P_n = P_n$, $d.b = b$, $d.c = c$, where $d$ is the tuple of $D$ that has been created in step 1, then stop, else go to next step.\\
4. If \( (user\_added,P_N, b, c, f, \mathit{P_x}) \xleftarrow{\text{\(\tau_2 \)}} \mathit{F_{bch}}\) and if $d.P_n \neq P_n \lor d.b \neq b \lor d.c \neq c$, where $d$ is the tuple of $D$ that has been created in step 1, then become inactive to prevent state update, else go to next step.\\
5. If \((user\_add\_timeout, P_N, b, c, f) \xleftarrow{\text{\(\tau_3 = \tau_0 + \mathit{n\Delta} \)}} \mathit{Z}\) then \\
\((refund, P_N, b, c, f) \xrightarrow{\text{\(\tau_3 \)}} \mathit{F_{bch}}\) and wait.\\
6. If \((no\_refund\_cancel, P_N, b, c, f) \xleftarrow{\text{\(\tau_4  = \tau_3 + \mathit{n\Delta}\)}} \mathit{Z}\) then \\
\( (withdraw,P_N, b, c, f) \xrightarrow{\text{\(\tau_4\)}} \mathit{F_{bch}}\)
\item Back to Party \(\mathit{P_x}\)\\
7. Upon \( (refund\_request,P_N, b, c, f) \xleftarrow{\text{\(\tau_5 = \tau_3 + \mathit{\Delta}\)}} \mathit{F_{bch}}\) \\
If \(\mathit{P_N}\in b.members\) then \\
\( (cancel\_refund,P_N, b, c, f, S_b) \xrightarrow{\text{\(\tau_5\)}} \mathit{F_{bch}}\) else stop.\\ 

\hrulefill

Functionality \(\mathit{F_{bch}}\) : adding a new participant

\hrulefill
\begin{itemize}
\item \textbf{Upon} \( (deposit,P_N, b, c, f) \xleftarrow{\text{\(\tau_0 \)}} \mathit{P}\)\\
1. If \(\mathit{P} \neq \mathit{P_N}\) or b channel does not exist or 
\(\mathit{P}\in b.members\) or \( f \leq 0\) or \( c \leq 0\) then stop.
else go to next step.\\
2. Store in deposit storage $D \in \Lambda$ the tuple $d:=(P_N, b, c, f, tau_0)$
\( (add\_request,P_N, b, c, f) \xleftarrow{\text{\(\tau_0 + \mathit{\Delta}\)}} b.members\)
\item \textbf{Upon} \( (add\_user,P_N, b, c, f) \xrightarrow{\text{\(\tau_2 \)}} \mathit{P_x}\)\\
1. If there is no tuple \( (P_N, b, c, f) \) in storage \(D \) or \(\tau_2 - \tau_0 > \mathit{n\Delta} \) then stop, else go to the next step.\\
2.  \(b.members = b.members \cup \{P_N\}\), \\
    \(b.balance[P_N] = c\), \\
    \(b.balance[P_x] = b.balance[P_x] + f \), \\
3. \( (user\_added,P_N, b, c, f, \mathit{P_x}) \xleftarrow{\text{\(\tau_2 + \mathit{\Delta} \)}} b.members\) 

\item \textbf{Upon} \((refund, P_N, b, c, f) \xleftarrow{\text{\(\tau_3 \)}} \mathit{P_N}\)\\
1. Store in refund storage \(R \in \Lambda \) the tuple \( r:=(P_N, b, c, f,\tau_3  ) \)\\
2. \( (refund\_request,P_N, b, c, f) \xrightarrow{\text{\(\tau_3 + \mathit{\Delta}\)}} b.members\) and wait.

\item \textbf{Upon} \( (cancel\_refund,P_N, b, c, f, S_b) \xleftarrow{\text{\(\tau_5 \)}} P_k\)\\
1. If \(\tau_5 \leq \tau_3 + n\Delta\) and if tuple r exists in R where \(r.P_N, r.b\) are equal to the attributes of cancel\_refund and \(S_b\) is valid and \( P_N \in  b.members\) then \\
remove tuple r from R storage else go to the next step.

2. \( (refund\_cancelled, P_N, b, c, f) \xrightarrow{\text{\(\tau_5 + \mathit{\Delta}\)}} b.members\). 
\end{itemize}

\end{itemize} 

\subsection{Security analysis}
\label{subsec:security}

In this section, we prove that the protocol \(\Pi\) emulates the ideal \(\mathit{F_{bch}}\) and we construct a simulator \(\mathit{Sim}\) for every adversary \(\mathit{Adv}\)  that operates in the \(\mathit{F_{bch}^ \mathit{L}}\)  - world. Due to the complexity of the origami ecosystem, a modular approach will facilitate the security analysis approach of our scheme. In section \ref{subsec:idealfunctionality} we presented the ideal functionality for the update action of adding a member to the base channel, which is the core of the origami model and can sufficiently describe the same function performed in the groups constructed upon itself, requiring only limited modifications. 

Before the description of the simulator Sim, we define some security guarantees that are regarded as essential for the origami construction model:
\begin{itemize}
\item \textbf{membership validity}
The actions executed within each instance of the Origami protocols should be operated only by members of that instance. No external member can interact with the instances of the protocol.
\item \textbf{balance security}
The amounts of coins presented within the instances of the protocols should correspond to the amount of coins deposited on-chain and be distributed to participants according to the valid state updates of the instances.
\item \textbf{Consensus on update}
As a state channel network, Origami is essential to guarantee the consent of all members to the change of state. An instance of the contract of a group can be successfully updated when all the members of the group have approved the transition. 
\end{itemize}
 We will separately discuss each part of the protocol in order to thoroughly examine all possible corrupted behaviours.  The simulator \(\mathit{Sim}\) presented below will run a copy of the hybrid world where it receives all input messages of the honest parties from the ideal functionality and the messages the adversary \(\mathit{Adv}\) 
gives to the hybrid ideal functionality. The purpose of this description is to prove the indistinguishability of the ideal and the hybrid world for every corruption combination.

\subsection{Adding participant}
\label{subsec:simulator}

Regarding the adding of a new participant, we describe three possible scenarios where the existing member \(\mathit{P_x}\) a) adds a user \(\mathit{P_N}\) without their consensus or b) alters the amounts of coins \(\mathit{c,f}\), that \(\mathit{P_N}\) locked in the previous step, as described in \ref{subsec:basechannel}, and c) where the new member \(\mathit{P_N}\) claims the refund of the deposit while being already accepted in the group. 

\begin{tcolorbox}[title= Simulator Sim, colback=white]

\textbf{\(\mathit{P_x}\) is corrupt} 
\newline
\newline
a)  Upon \( (P_x,add\_user,P_N, b, c, f) \xleftarrow{}\mathit{F_{bch}}\)\\
If there is no tuple \( (P_N, b, c, f) \) in storage \(D \) then stop.\\

b) Upon \( (P_x,add\_user,P_N, b, c, f) \xleftarrow{\text{\(\tau\)}}\mathit{F_{bch}}\)\\
1. If there is a tuple \( (P_N, b, c, f) \) stored at \(\tau_0 \) in storage \(D \) and \(\tau < \tau_0 + \mathit{n\Delta} \) then add user on behalf of \(\mathit{P_x}\) and  send to all b.members \( (user\_added,P_N, b, c, f, \mathit{P_x}) \xleftarrow{\text{\(\tau_2 + \mathit{\Delta} \)}} \mathit{F_{bch}}\) \\
2. if $d.P_n \neq P_n \lor d.b \neq b \lor d.c \neq c$, where $d$ is the tuple of $D$ that has been stored at \(\tau_0 \) in storage \(D \) then \(\mathit{P_N}\) becomes inactive to prevent state update.\\

\textbf{\(\mathit{P_N}\) is corrupt\\} \\
c)  Upon \( (refund\_request, PN , b, c, f ) \xleftarrow{}\mathit{F_{bch}}\)\\
If \(\mathit{P_N}\in b.members\) then \\
\( (cancel\_refund,P_N, b, c, f, S_b) \xleftarrow{} \mathit{F_{bch}}\)
\end{tcolorbox}

\section{Conclusions}

State channels, like other scalability mechanisms, are a relatively novel addition to an already recent technology. It is therefore possible to evaluate their effect and progress since their introduction. The conclusion that stems from this analysis \cite{negka2021blockchain} is that in a strictly defined environment, state channels are quite beneficial. However, they have mostly failed so far to broaden their scope enough to make significant impact. Limited applicability, as derived from the numerous limitations visible in \ref{table:comparison} prevents this approach from applying its full potential in combating the scalability issue.

Most layer 2 techniques aim to increase the transaction processing capacity of the blockchain network. State channels instead bring down the overall need for on-chain transactions and as a result could be very beneficial in conjunction with other, more prevailing, layer 2 approaches.

With that in mind, it was deemed beneficial to focus on the development of a state channel scheme that could circumvent the existing obstacles and enable the reaping of the full benefits of the technology. Origami is meant to fill what has been missing from this research field, a design with a better balance between the absence of requirements and the presence of features. 

It can be confidently concluded that the goal has been achieved. Through Origami, applications can benefit from state channels regardless of whether they employ turn-based communication schemes or not. Unnecessary termination of channels and therefore the necessity to interact with the blockchain to set new ones up have been eliminated. Users can join a boundlessly scaling ecosystem that reduces on-chain transactions. Origami, in the optimistic case, requires a single on-chain interaction per person indefinitely, as opposed to existing designs that require setting up a new state channel for every use or alteration of the participant set.

The Origami design even prioritises keeping the storage requirements to a minimum by preventing the state from endlessly expanding and burdening the user nodes. Future work intentions include the development of a built-in mechanism that bypasses the constant connectivity requirement without relying upon an external service, as well as a high-level extensive security analysis for the Origami design.
\label{sec:conclusion}


\bibliographystyle{unsrt}  
\bibliography{paper}

\begin{thebibliography}{10}

\bibitem{treiblmaier2020blockchain}
Horst Treiblmaier and Trevor Clohessy.
\newblock {\em Blockchain and Distributed Ledger Technology Use Cases:
  Applications and Lessons Learned}.
\newblock 01 2020.

\bibitem{nakamoto2019bitcoin}
Satoshi Nakamoto.
\newblock Bitcoin: A peer-to-peer electronic cash system.
\newblock Technical report, Manubot, 2019.

\bibitem{buterin2014next}
Vitalik Buterin et~al.
\newblock A next-generation smart contract and decentralized application
  platform.
\newblock {\em white paper}, 3(37), 2014.

\bibitem{khan2021systematic}
Dodo Khan, Low~Tang Jung, and Manzoor~Ahmed Hashmani.
\newblock Systematic literature review of challenges in blockchain scalability.
\newblock {\em Applied Sciences}, 11(20):9372, 2021.

\bibitem{gervais2016security}
Arthur Gervais, Ghassan~O Karame, Karl W{\"u}st, Vasileios Glykantzis, Hubert
  Ritzdorf, and Srdjan Capkun.
\newblock On the security and performance of proof of work blockchains.
\newblock In {\em Proceedings of the 2016 ACM SIGSAC conference on computer and
  communications security}, pages 3--16, 2016.

\bibitem{saleh2021blockchain}
Fahad Saleh.
\newblock Blockchain without waste: Proof-of-stake.
\newblock {\em The Review of financial studies}, 34(3):1156--1190, 2021.

\bibitem{zhou2020solutions}
Qiheng Zhou, Huawei Huang, Zibin Zheng, and Jing Bian.
\newblock Solutions to scalability of blockchain: A survey.
\newblock {\em IEEE Access}, 8:16440--16455, 2020.

\bibitem{bach2018comparative}
Leo~Maxim Bach, Branko Mihaljevic, and Mario Zagar.
\newblock Comparative analysis of blockchain consensus algorithms.
\newblock In {\em 2018 41st International Convention on Information and
  Communication Technology, Electronics and Microelectronics (MIPRO)}, pages
  1545--1550. Ieee, 2018.

\bibitem{chow2018sharding}
Sherman~SM Chow, Ziliang Lai, Chris Liu, Eric Lo, and Yongjun Zhao.
\newblock Sharding blockchain.
\newblock In {\em 2018 IEEE International Conference on Internet of Things
  (iThings) and IEEE Green Computing and Communications (GreenCom) and IEEE
  Cyber, Physical and Social Computing (CPSCom) and IEEE Smart Data
  (SmartData)}, pages 1665--1665. IEEE, 2018.

\bibitem{kokoris2018omniledger}
Eleftherios Kokoris-Kogias, Philipp Jovanovic, Linus Gasser, Nicolas Gailly,
  Ewa Syta, and Bryan Ford.
\newblock Omniledger: A secure, scale-out, decentralized ledger via sharding.
\newblock In {\em 2018 IEEE Symposium on Security and Privacy (SP)}, pages
  583--598. IEEE, 2018.

\bibitem{gangwal2023survey}
Ankit Gangwal, Haripriya~Ravali Gangavalli, and Apoorva Thirupathi.
\newblock A survey of layer-two blockchain protocols.
\newblock {\em Journal of Network and Computer Applications}, 209:103539, 2023.

\bibitem{green2017bolt}
Matthew Green and Ian Miers.
\newblock Bolt: Anonymous payment channels for decentralized currencies.
\newblock In {\em Proceedings of the 2017 ACM SIGSAC Conference on Computer and
  Communications Security}, pages 473--489, 2017.

\bibitem{avarikioti2018towards}
Georgia Avarikioti, Felix Laufenberg, Jakub Sliwinski, Yuyi Wang, and Roger
  Wattenhofer.
\newblock Towards secure and efficient payment channels.
\newblock {\em arXiv preprint arXiv:1811.12740}, 2018.

\bibitem{negka2021blockchain}
Lydia~D Negka and Georgios~P Spathoulas.
\newblock Blockchain state channels: A state of the art.
\newblock {\em IEEE Access}, 2021.

\bibitem{thibault2022blockchain}
Louis~Tremblay Thibault, Tom Sarry, and Abdelhakim~Senhaji Hafid.
\newblock Blockchain scaling using rollups: A comprehensive survey.
\newblock {\em IEEE Access}, 2022.

\bibitem{back2014enabling}
Adam Back, Matt Corallo, Luke Dashjr, Mark Friedenbach, Gregory Maxwell, Andrew
  Miller, Andrew Poelstra, Jorge Tim{\'o}n, and Pieter Wuille.
\newblock Enabling blockchain innovations with pegged sidechains.
\newblock {\em URL: http://www. opensciencereview.
  com/papers/123/enablingblockchain-innovations-with-pegged-sidechains}, 72,
  2014.

\bibitem{singh2020sidechain}
Amritraj Singh, Kelly Click, Reza~M Parizi, Qi~Zhang, Ali Dehghantanha, and
  Kim-Kwang~Raymond Choo.
\newblock Sidechain technologies in blockchain networks: An examination and
  state-of-the-art review.
\newblock {\em Journal of Network and Computer Applications}, 149:102471, 2020.

\bibitem{poon2017plasma}
Joseph Poon and Vitalik Buterin.
\newblock Plasma: Scalable autonomous smart contracts.
\newblock {\em White paper}, pages 1--47, 2017.

\bibitem{coleman2018counterfactual}
Jeff Coleman, Liam Horne, and Li~Xuanji.
\newblock Counterfactual: Generalized state channels.
\newblock {\em Acessed: Nov}, 4:2019, 2018.

\bibitem{close2018forcemove}
Tom Close and Andrew Stewart.
\newblock Forcemove: An n-party state channel protocol, 2018.

\bibitem{buckland2019two}
Chris Buckland and Patrick McCorry.
\newblock Two-party state channels with assertions.
\newblock In {\em Financial Cryptography Workshops}, pages 3--11, 2019.

\bibitem{chakravarty2020hydra}
Manuel~MT Chakravarty, Sandro Coretti, Matthias Fitzi, Peter Gazi, Philipp
  Kant, Aggelos Kiayias, and Alexander Russell.
\newblock Hydra: Fast isomorphic state channels.
\newblock {\em IACR Cryptol. ePrint Arch.}, 2020:299, 2020.

\bibitem{miller2020you}
Andrew Miller.
\newblock You sank my battleship! a case study to evaluate state channels as a
  scaling solution for cryptocurrencies.
\newblock In {\em Financial Cryptography and Data Security: FC 2019
  International Workshops, VOTING and WTSC, St. Kitts, St. Kitts and Nevis,
  February 18--22, 2019, Revised Selected Papers}, volume 11599, page~35.
  Springer Nature, 2020.

\bibitem{dziembowski2017perun}
Stefan Dziembowski, Lisa Eckey, Sebastian Faust, and Daniel Malinowski.
\newblock Perun: Virtual payment channels over cryptographic currencies.
\newblock {\em IACR Cryptol. ePrint Arch.}, 2017:635, 2017.

\bibitem{miller2019sprites}
Andrew Miller, Iddo Bentov, Surya Bakshi, Ranjit Kumaresan, and Patrick
  McCorry.
\newblock Sprites and state channels: Payment networks that go faster than
  lightning.
\newblock In {\em International Conference on Financial Cryptography and Data
  Security}, pages 508--526. Springer, 2019.

\bibitem{dziembowski2018general}
Stefan Dziembowski, Sebastian Faust, and Kristina Host{\'a}kov{\'a}.
\newblock General state channel networks.
\newblock In {\em Proceedings of the 2018 ACM SIGSAC Conference on Computer and
  Communications Security}, pages 949--966, 2018.

\bibitem{dziembowski2019multi}
Stefan Dziembowski, Lisa Eckey, Sebastian Faust, Julia Hesse, and Kristina
  Host{\'a}kov{\'a}.
\newblock Multi-party virtual state channels.
\newblock In {\em Annual International Conference on the Theory and
  Applications of Cryptographic Techniques}, pages 625--656. Springer, 2019.

\bibitem{close2019nitro}
Tom Close.
\newblock Nitro protocol.
\newblock {\em IACR Cryptol. ePrint Arch.}, 2019:219, 2019.

\bibitem{mccorry2019pisa}
Patrick McCorry, Surya Bakshi, Iddo Bentov, Sarah Meiklejohn, and Andrew
  Miller.
\newblock Pisa: Arbitration outsourcing for state channels.
\newblock In {\em Proceedings of the 1st ACM Conference on Advances in
  Financial Technologies}, pages 16--30, 2019.

\bibitem{avarikioti2019brick}
Georgia Avarikioti, Eleftherios~Kokoris Kogias, Roger Wattenhofer, and Dionysis
  Zindros.
\newblock Brick: Asynchronous payment channels.
\newblock {\em arXiv preprint arXiv:1905.11360}, 2019.

\bibitem{kumar2014}
Amrit Kumar, Pascal Lafourcade, and Cédric Lauradoux.
\newblock Performances of cryptographic accumulators.
\newblock In {\em 39th Annual IEEE Conference on Local Computer Networks},
  pages 366--369, 2014.

\bibitem{ozcelik2021overview}
Ilker Ozcelik, Sai Medury, Justin Broaddus, and Anthony Skjellum.
\newblock An overview of cryptographic accumulators.
\newblock {\em arXiv preprint arXiv:2103.04330}, 2021.

\bibitem{Boneh}
Dan Boneh, Benedikt B{\"u}nz, and Ben Fisch.
\newblock Batching techniques for accumulators with applications to iops and
  stateless blockchains.
\newblock In Alexandra Boldyreva and Daniele Micciancio, editors, {\em Advances
  in Cryptology -- CRYPTO 2019}, pages 561--586, Cham, 2019. Springer
  International Publishing.

\bibitem{Lipmaa2012SecureAF}
Helger Lipmaa.
\newblock Secure accumulators from euclidean rings without trusted setup.
\newblock In {\em ACNS}, 2012.

\bibitem{Wesolowski}
Benjamin Wesolowski.
\newblock Efficient verifiable delay functions.
\newblock 11478:379--407, 2019.

\bibitem{Dobson}
Samuel Dobson, Steven Galbraith, and Benjamin Smith.
\newblock Trustless unknown-order groups.
\newblock {\em Mathematical Cryptology}, 1(2):25–39, Mar. 2022.

\bibitem{NEGKA2023100114}
Lydia Negka, Angeliki Katsika, Georgios Spathoulas, and Vassilis Plagianakos.
\newblock Blockchain state channels with compact states through the use of rsa
  accumulators.
\newblock {\em Blockchain: Research and Applications}, 4(1):100114, 2023.

\bibitem{Baldimtsi2017}
Foteini Baldimtsi, Jan Camenisch, Maria Dubovitskaya, Anna Lysyanskaya, Leonid
  Reyzin, Kai Samelin, and Sophia Yakoubov.
\newblock Accumulators with applications to anonymity-preserving revocation.
\newblock Cryptology ePrint Archive, Report 2017/043, 2017.
\newblock \url{https://ia.cr/2017/043}.

\bibitem{canetti2005security}
R~Canetti.
\newblock Security, uc: A new paradigm for cryptographic protocols.
\newblock In {\em Proceedings of the 42nd Annual Symposium on Foundations of
  Computer Science FOCS}, 2005.

\end{thebibliography}

\end{document}